\documentclass{revtex4}
\usepackage{graphicx}

\newcommand{\be}{\begin{equation}}
\newcommand{\ee}{\end{equation}}
\newcommand{\ba}{\begin{eqnarray}}
\newcommand{\ea}{\end{eqnarray}}
\newcommand{\ban}{\begin{eqnarray*}}
\newcommand{\ean}{\end{eqnarray*}}

\begin{document}

\title{Chromodynamic Fluctuations in Quark-Gluon Plasma}

\author{Stanis\l aw Mr\' owczy\' nski\footnote{Electronic address:
{\tt mrow@fuw.edu.pl}}}

\affiliation{ Institute of Physics, \'Swi\c etokrzyska Academy \\
ul.~\'Swi\c etokrzyska 15, PL - 25-406 Kielce, Poland \\
and So\l tan Institute for Nuclear Studies \\
ul.~Ho\.za 69, PL - 00-681 Warsaw, Poland}

\date{4-th April 2008}

\begin{abstract} 

Fluctuations of chromodynamic fields in the collisionless quark-gluon 
plasma are found as a solution of the initial value linearized problem. 
The plasma initial state is on average colorless, stationary and 
homogeneous. When the state is stable, the initial fluctuations 
decay exponentially and in the long-time limit a stationary spectrum 
of fluctuations is established. For the equilibrium plasma it reproduces 
the spectrum which is provided by the fluctuation-dissipation relation. 
Fluctuations in the unstable plasma, where the memory of initial 
fluctuations is not lost, are also discussed.

\end{abstract}

\pacs{PACS: 12.38.Mh, 05.20.Dd, 11.10.Wx}


\maketitle


\section{Introduction}


In the quark-gluon plasma (QGP), which is on average locally colorless,
chromodynamic fields, color charges and currents experience random 
fluctuations which appear to influence dynamics of the whole system. 
In the equilibrium plasmas there are characteristic stationary spectra 
of fluctuations which can be found by means of the fluctuation-dissipation
relations. Fluctuation spectra in nonequilibrium systems evolve in time, 
and their characteristics usually depend on an initial state of the 
system. Our aim is to develop a method to study chromodynamic fluctuations 
in equilibrium and nonequilibrium QGP. We are particularly interested in 
QGP produced at the early stage of relativistic heavy-ion collisions. 
Such a plasma is presumably unstable with respect to chromomagnetic modes 
due to anisotropic momentum distribution of quarks and gluons (partons),  
see the review \cite{Mrowczynski:2005ki}. The instability growth is 
associated with generation of chromomagnetic fields which in turn 
influence various plasma properties. Transport coefficients of 
such a plasma, which are controlled by the fluctuation spectrum of 
chromomagnetic fields, are then strongly modified \cite{Asakawa:2006jn}. 
The fluctuation spectra can be obtained from numerical simulations 
of the unstable QGP, which have been successfully performed by several groups
\cite{Arnold:2005vb,Arnold:2005ef,Rebhan:2005re,Dumitru:2006pz,Romatschke:2006nk,Bodeker:2007fw,Berges:2007re}, 
but the problem has not attracted much attention yet; there are no 
analytical studies as well. 

Fluctuations can be theoretically described using several methods 
reviewed in the classical monographs \cite{Akh75,Sit82}. Field-theory 
techniques developed for relativistic equilibrium plasmas are worked
out in \cite{Siv85,Lemoine:1995fh}. The method, which seems to be 
physically most appealing, is clearly exposed in the handbook \cite{LP81}. 
The method - applicable to both equilibrium and nonequilibrium 
plasmas - provides the spectrum of fluctuations as a solution of the 
initial value (linearized) problem. The initial plasma state is assumed 
to be on average charge neutral, stationary and homogeneous. When the 
state is stable, the initial fluctuations are explicitly shown to 
exponentially decay and in the long time limit one finds a stationary 
spectrum of fluctuations. In this way one obtains for the equilibrium 
plasma the spectrum which is alternatively provided by the 
fluctuation-dissipation relation. When the initial state is unstable, 
the memory of initial fluctuations is not lost, as the unstable modes, 
which are present in the initial fluctuation spectrum, exponentially 
grow. We apply the method to study chromodynamic fluctuations in the 
quark-gluon plasma. 

The analysis presented here closely follows our paper \cite{Mrow2007} 
where electromagnetic fluctuations in the electron-ion plasma are discussed. 
Since the fluctuation spectra are found as solutions of linearized equations 
of motion, our chromodynamic and electromagnetic considerations are quite 
similar to each other. However, some points significantly differ. First 
of all, the starting point is different; the nonAbelian equations become 
Abelian only after the linearization. Solutions of linearized nonAbelian 
equations are usually gauge noncovariant and should be modified to comply 
with the gauge covariance. The correlation functions derived here in the 
linear approximation appear to be, as discussed in Sec.~\ref{subsec-gauge}, 
gauge invariant but the result is a priori is not evident. Color charges 
are of different nature than electric ones and need an adequate treatment. 
Therefore, the chromodynamic results cannot be trivially inferred from their 
electromagnetic counterparts.

An approach to fluctuations similar to the one adopted here is sometimes 
called the Klimontovich method. It was earlier used by Litim and Manuel 
\cite{Litim:1999id,Litim:2001db} to derive collisions terms of the transport 
equations of quark-gluon plasma close to equilibrium. These authors, however, 
treated color charges of partons as classical variables while in our study 
the color degrees of freedom are of quantum mechanical nature. 

Our paper is organized as follows. In Sec.~\ref{sec-preliminaries} we 
present the theoretical framework to be used in our further considerations. 
The linearized kinetic equation are solved together with Maxwell equations 
by means of the one-sided Fourier transformation in 
Sec.~\ref{sec-initial-value}. Fluctuation spectra of chromodynamic 
fields are expressed through the initial fluctuations which are calculated
in Sec.~\ref{sec-initial-fluc}. The initial fluctuations are identified 
with fluctuations in the systems of free quarks and gluons. Fluctuations 
of chromomagnetic and chromoelectric fields in the stable isotropic plasma 
are discussed in, respectively, Sec.~\ref{sec-fluc-B} and \ref{sec-fluc-E},
while longitudinal electric fields in the unstable two-stream system are 
studied in Sec.~\ref{sec-2-streams}. In the last case, the fluctuation 
spectrum strongly depends on the initial state. We close the paper with 
Sec.~\ref{sec-discussion} where our results are extensively discussed. 
In particular, a gauge independence of the correlation functions of 
interest is demonstrated. We also mention how to compute fluctuations 
of charges and currents which are not analyzed in the main part of our
paper. The Appendix presents a field-theory derivation of the correlation 
function of distribution functions of free but colored partons. 

Throughout the article we use the natural units with $c=k_B = 1$
and four types of indices: $m,n, \dots$ and $a,b, \dots$ label, 
respectively, color components in the fundamental and adjoint 
 representation of ${\rm SU}(N_c)$ gauge group; the indices 
$\mu,\nu, \dots$ and $i,j, \dots$ are used to label four- and 
three-vectors, respectively. In the Appendix the indices $i,j$
denote internal degrees of freedom of scalar fields.


\section{Preliminaries}
\label{sec-preliminaries}


The transport theory of weakly coupled quark-gluon plasma, which 
forms the basis  of our analysis, is formulated in terms particles and 
classical fields. The particles - quarks, antiquarks and gluons - 
should be understood as sufficiently hard quasiparticle excitations of
quantum fields of QCD while the classical fields are highly 
populated soft gluonic modes. An excitation is called `hard', when 
its momentum in the equilibrium system is of order of temperature 
$T$, and it is called `soft' when the momentum is $gT$ with $g$
being the coupling constant which is assumed to be small. In our 
further considerations the quasiparticles are treated as classical 
particles obeying Boltzmann statistics but, as shown in 
Sec.~\ref{subsec-statistics}, the effect of quantum statistics can 
be easily taken into account.

The transport equations of quarks, antiquarks and gluons read
\ba
\nonumber
\big(D^0 + {\bf v} \cdot {\bf D} \big) Q(t,{\bf r},{\bf p})
- {g \over 2}
\{{\bf E}(t,{\bf r}) + {\bf v} \times {\bf B}(t,{\bf r}), 
\nabla_p Q(t,{\bf r},{\bf p}) \}
&=& 0 \;, \\ [2mm]
\label{transport-eq}
\big(D^0 + {\bf v} \cdot {\bf D} \big) \bar Q(t,{\bf r},{\bf p})
+ {g \over 2}
\{{\bf E}(t,{\bf r}) + {\bf v} \times {\bf B}(t,{\bf r}), 
\nabla_p \bar Q(t,{\bf r},{\bf p}) \}
&=& 0 \;, \\ [2mm]
\nonumber
\big(D^0 + {\bf v} \cdot {\bf D} \big) G(t,{\bf r},{\bf p})
- {g \over 2}
\{{\bf E}(t,{\bf r}) + {\bf v} \times {\bf B}(t,{\bf r}), 
\nabla_p G(t,{\bf r},{\bf p}) \}
&=&  0\;.
\ea
The (anti-)quark distribution functions $Q(t,{\bf r},{\bf p})$ 
and $\bar Q(t,{\bf r},{\bf p})$, which are $N_c\times N_c$ hermitean 
matrices, belong to the fundamental representation of the SU($N_c$) 
group, while the gluon distribution function $G(t,{\bf r},{\bf p})$, 
which is a $(N_c^2 -1) \times (N_c^2 -1)$ matrix, belongs to the adjoint 
representation. The distribution functions depend on the time ($t$), 
position (${\bf r}$) and momentum (${\bf p}$) variables. There is
no dependence on 0-th component of the four-vector $p^\mu$ as the 
distribution functions are assumed to be non-zero only for momenta
obeying the mass-shell constraint $p^\mu p_\mu =0$. Because the 
partons are assumed to be massless, the velocity ${\bf v}$ equals 
${\bf p}/E_{\bf p}$ with $E_{\bf p}=|{\bf p}|$. The covariant 
derivative, which in the four-vector notation reads 
$D^\mu \equiv \partial^\mu - ig [A^\mu(x),\cdots \;]$, as well as
the chromodynamic fields ${\bf E}(t,{\bf r})$ and ${\bf B}(t,{\bf r})$ 
belong to either the fundamental or adjoint representation, 
correspondingly. To simplify the notation we use the same symbols 
$D^0$, ${\bf D}$, ${\bf E}$ and ${\bf B}$ to denote a given quantity 
in the fundamental or adjoint representation. The symbol 
$\{\dots , \dots \}$ denotes the anticommutator. Since the 
fluctuations of interest are assumed to be of the time scale, which 
is much shorter than that of inter-parton collisions, the collision 
terms are absent in Eqs.~(\ref{transport-eq}). The approximation is
further discussed in Sec.~\ref{subsec-linear}.

The transport equations are supplemented by the nonAbelian version
of Maxwell equations describing a self-consistent generation of 
the chromoelectric and chromomagnetic fields. The equations read
\ba
\nonumber
{\bf D} \cdot {\bf E}(t, {\bf r}) &=&  \rho (t, {\bf r}) 
\;,\;\;\;\;\;\;\;\;\;\;\;\;\;
{\bf D} \cdot {\bf B}(t, {\bf r}) = 0 \;, 
\\[2mm]
\label{Maxwell-eqs-x}
{\bf D} \times {\bf E}(t, {\bf r}) &=& - 
D_0 {\bf B}(t, {\bf r}) \;,
\;\;\;\;\;
{\bf D} \times {\bf B}(t, {\bf r}) = 
{\bf j}(t, {\bf r}) + D_0 {\bf E}(t, {\bf r}) \;,
\ea
where the color four-current $j^\mu =(\rho , {\bf j})$ in the adjoint
representation equals
\be
\label{current-def}
j^\mu_a (t,{\bf r}) = - g \int {d^3 p \over (2\pi)^3} \,
\frac{p^\mu}{E_{\bf p}}
{\rm Tr}\Big[\tau^a \big(Q (t,{\bf r},{\bf p}) 
- \bar Q(t,{\bf r},{\bf p}) \big)
+ T^a G(t,{\bf r},{\bf p}) \Big] \;,
\ee
where $\tau^a$, $T^a$ with $a = 1, ... \, ,N_c^2-1$ are the SU($N_c$)
group generators in the fundamental and adjoint representations,
normalized as ${\rm Tr}[\tau^a \tau^b] = \frac12 \delta^{ab}$ and
${\rm Tr}[T^a T^b] = N_c \delta^{ab}$. The set of transport 
(\ref{transport-eq}) and Maxwell (\ref{Maxwell-eqs-x}) equations 
is covariant with respect to ${\rm SU}(N_c)$ gauge transformations. 

We are going to consider small deviations from a stationary homogeneous 
state described by $Q^0({\bf p})$, $\bar Q^0({\bf p})$ and $G^0({\bf p})$.
The state is globally and locally colorless; there are no currents as well. 
Therefore,
\be
Q^0_{nm}({\bf p}) = n ({\bf p}) \: \delta^{nm} \;,
\;\;\;\;\;\;\;
\bar Q^0_{nm}({\bf p}) = \bar n ({\bf p}) \: \delta^{nm} \;,
\;\;\;\;\;\;\;
G^0_{ab}({\bf p}) = n_g ({\bf p}) \: \delta^{ab} \;.
\ee
The indices $n,m,\dots$ and $a,b,\dots$ refer, as already mentioned, 
to the fundamental and adjoint representation, respectively. Due to 
the absence of color charges and currents in the stationary and 
homogeneous state, the chromoelectric ${\bf E}(t,{\bf r})$ and 
chromomagnetic ${\bf B}(t,{\bf r})$ fields are expected to vanish
while the potentials $A^0(t,{\bf r}), {\bf A}(t,{\bf r})$ are of 
pure gauge only. Since the plasma under considerations is assumed
to be weakly coupled with the perturbative vacuum state, the 
potentials can be gauge away to vanish. 

We write down the quark distribution function as
$Q(t,{\bf r},{\bf p}) =  Q^0({\bf p}) + 
\delta Q(t,{\bf r},{\bf p})$, and we assume that
\be
\label{q-conditions}
|Q^0| \gg |\delta Q| \;, \;\;\;\;\;\;\;\;
|\nabla_p Q^0| \gg | \nabla_p \delta Q| \;,
\ee
with the analogous formulas for antiquarks and gluons.
We linearize the transport (\ref{transport-eq}) and Maxwell 
(\ref{Maxwell-eqs-x}) equations in the deviations from the 
stationary homogeneous state. We assume that $\delta Q$, 
$\delta \bar Q$, $\delta G$, ${\bf E}$, ${\bf B}$, $A^0$ and 
${\bf A}$ are all of the same order. Validity of the 
approximation is further discussed in Sec.~\ref{subsec-linear}.
The linearized transport equations are
\ba
\nonumber 
\Big(\frac{\partial}{\partial t} + {\bf v} \cdot \nabla \Big) 
\delta Q(t,{\bf r},{\bf p})
- g \big({\bf E}(t,{\bf r}) + {\bf v} \times {\bf B}(t,{\bf r})\big) 
\nabla_p n({\bf p})
&=& 0 \;, 
\\ [2mm] \label{trans-eq-lin}
\Big(\frac{\partial}{\partial t} + {\bf v} \cdot \nabla \Big) 
\delta \bar Q(t,{\bf r},{\bf p})
+ g \big({\bf E}(t,{\bf r}) + {\bf v} \times {\bf B}(t,{\bf r})\big) 
\nabla_p \bar n({\bf p})
&=& 0 \;, \\ [2mm]
\nonumber
\Big(\frac{\partial}{\partial t} + {\bf v} \cdot \nabla \Big) 
\delta G(t,{\bf r},{\bf p})
- g \big({\bf E}(t,{\bf r}) + {\bf v} \times {\bf B}(t,{\bf r})\big) 
\nabla_p n_g({\bf p})
&=&  0\;,
\ea
while the Maxwell equations get the form familiar from the 
electrodynamics (in the so-called Heaviside-Lorentz system of units)
\ba
\label{Maxwell-eqs-x-lin}
\nabla \cdot {\bf E}(t, {\bf r}) &=& \rho (t, {\bf r}) 
\;,\;\;\;\;\;\;\;\;\;\;\;\;\;
\nabla \cdot {\bf B}(t, {\bf r}) = 0 \;, \\[2mm]
\nonumber
\nabla \times {\bf E}(t, {\bf r}) &=& - 
{\partial {\bf B}(t, {\bf r}) \over \partial t} \;,
\;\;\;\;\;
\nabla \times {\bf B}(t, {\bf r}) = 
{\bf j}(t, {\bf r}) +{\partial {\bf E}(t, {\bf r}) \over \partial t} \;,
\ea
with
\be
\label{current-lin}
j^\mu_a (t,{\bf r}) = - g \int {d^3 p \over (2\pi)^3} \,
\frac{p^\mu}{E_{\bf p}} \;\delta N_a(t,{\bf r},{\bf p})
\;,
\ee
where 
\be
\label{N-def}
\delta N_a(t,{\bf r},{\bf p}) \equiv
{\rm Tr}\big[\tau^a \big(\delta Q (t,{\bf r},{\bf p}) 
- \delta \bar Q(t,{\bf r},{\bf p}) \big)
+ T^a \delta G(t,{\bf r},{\bf p}) \big]\;.
\ee

The linearized equations are Abelian and they correspond to the 
multi-component electrodynamics of $N_c$ charges. It should be 
noted, however, that the gluon contribution to the color current, 
which is of obviously nonAbelian origin, is taken into account. 
The equations are no longer covariant with respect to ${\rm SU}(N_c)$ 
gauge transformations. The gauge independence of our final results 
will be demonstrated in Sec.~\ref{subsec-gauge}.


\section{Initial value problem}
\label{sec-initial-value}


We are going to solve the linearized transport 
(\ref{trans-eq-lin}) and Maxwell (\ref{Maxwell-eqs-x-lin}) equations  
with the initial conditions
\ban
\nonumber
\delta Q(t=0,{\bf r},{\bf p}) &=& \delta Q_0({\bf r},{\bf p})
\;,\;\;\;\;\;
\delta \bar Q(t=0,{\bf r},{\bf p}) = \delta \bar Q_0({\bf r},{\bf p})
\;,\;\;\;\;\;
\delta G(t=0,{\bf r},{\bf p}) = \delta G_0({\bf r},{\bf p})
\;,
\\[2mm]
\nonumber 
{\bf E}(t=0,{\bf r}) &=& {\bf E}_0({\bf r})
\;,\;\;\;\;\;
{\bf B}(t=0,{\bf r})= {\bf B}_0({\bf r}) \;.
\ean
We apply to the equations the one-sided Fourier transformation
defined as
\be
f(\omega,{\bf k}) = \int_0^\infty dt \int d^3r 
e^{i(\omega t - {\bf k}\cdot {\bf r})}
f(t,{\bf r}) \;.
\ee
The inverse transformation is 
\be
f(t,{\bf r}) = \int_{-\infty +i\sigma}^{\infty +i\sigma}
{d\omega \over 2\pi} \int {d^3k \over (2\pi)^3} 
e^{-i(\omega t - {\bf k}\cdot {\bf r})} f(\omega,{\bf k}) \;,
\ee
where the real parameter $\sigma > 0$ is chosen is such a
way that the integral over $\omega$ is taken along a straight
line in the complex $\omega-$plane, parallel to the real
axis, above all singularities of $f(\omega,{\bf k})$.

We note that 
\be
\int_0^\infty dt \int d^3r 
e^{i(\omega t - {\bf k}\cdot {\bf r})}
{\partial f(t,{\bf r}) \over \partial t}
= -i\omega f(\omega,{\bf k}) - f(t=0,{\bf k}) \;.
\ee

The linearized transport (\ref{trans-eq-lin}) and Maxwell 
equations (\ref{Maxwell-eqs-x-lin}), which are transformed by means 
of the one-sided Fourier transformation, read
\ba
\nonumber
-i(\omega - {\bf k}\cdot {\bf v}) \delta Q(\omega,{\bf k},{\bf p})
- g \big({\bf E}(\omega,{\bf k}) 
+ {\bf v} \times {\bf B}(\omega,{\bf k})\big)\cdot 
\nabla_p n({\bf p}) &=& \delta Q_0({\bf k},{\bf p})\;,
\\[2mm]
-i(\omega - {\bf k}\cdot {\bf v}) \delta \bar Q(\omega,{\bf k},{\bf p})
+ g \big({\bf E}(\omega,{\bf k}) 
+ {\bf v} \times {\bf B}(\omega,{\bf k})\big)\cdot 
\nabla_p \bar n({\bf p}) &=& \delta \bar Q_0({\bf k},{\bf p})\;,
\\[2mm] \nonumber
-i(\omega - {\bf k}\cdot {\bf v}) \delta G(\omega,{\bf k},{\bf p})
- g \big({\bf E}(\omega,{\bf k}) 
+ {\bf v} \times {\bf B}(\omega,{\bf k})\big)\cdot 
\nabla_p n_g({\bf p}) &=& \delta G_0({\bf k},{\bf p})\;,
\ea
\ba
\label{Maxwell-eqs-k}
i {\bf k} \cdot {\bf E}(\omega,{\bf k}) 
= \rho (\omega,{\bf k}) 
\;,\;\;\;\;\;\;\;\;\;\;\;\;\;
i {\bf k} \cdot {\bf B}(\omega,{\bf k}) &=& 0 \;, 
\\[2mm] \nonumber
i {\bf k} \times {\bf E}(\omega,{\bf k}) 
= i\omega {\bf B}(\omega,{\bf k}) + {\bf B}_0({\bf k}) \;,
\;\;\;\;\;
i {\bf k} \times {\bf B}(\omega,{\bf k}) 
&=& 
{\bf j}(\omega,{\bf k}) 
-i\omega {\bf E}(\omega,{\bf k}) - {\bf E}_0({\bf k}) \;.
\ea

One solves the transport equation as
\ba
\nonumber
\delta Q(\omega,{\bf k},{\bf p})
&=& \frac{i}{\omega - {\bf k}\cdot {\bf v}}
\Big( g \big({\bf E}(\omega,{\bf k})
+ {\bf v} \times {\bf B}(\omega,{\bf k})\big)\cdot
\nabla_p n({\bf p}) +  \delta Q_0({\bf k},{\bf p})\Big) \;,
\\[2mm] 
\label{solution1}
\delta \bar Q(\omega,{\bf k},{\bf p})
&=& \frac{i}{\omega - {\bf k}\cdot {\bf v}}
\Big( - g \big({\bf E}(\omega,{\bf k})
+ {\bf v} \times {\bf B}(\omega,{\bf k})\big)\cdot
\nabla_p \bar n ({\bf p}) +  \delta \bar Q_0({\bf k},{\bf p})\Big) \;,
\\[2mm] \nonumber
\delta G(\omega,{\bf k},{\bf p})
&=& \frac{i}{\omega - {\bf k}\cdot {\bf v}}
\Big( g \big({\bf E}(\omega,{\bf k})
+ {\bf v} \times {\bf B}(\omega,{\bf k})\big)\cdot
\nabla_p n_g({\bf p}) +  \delta G_0({\bf k},{\bf p})\Big) \;.
\ea

\subsection{Chromoelectric field}

Substituting the solutions (\ref{solution1}) into the Fourier
transformed current (\ref{current-lin}) and using the third Maxwell 
equation (\ref{Maxwell-eqs-k}) to express the magnetic field 
through the electric one, the current gets the form
\ba
{\bf j}_a(\omega,{\bf k}) &=& - i\frac{g^2}{2} \int {d^3p \over (2\pi)^3} \,
\frac{{\bf v}}{\omega - {\bf v}\cdot {\bf k}}
\Big(\big(1-\frac{{\bf k}\cdot {\bf v}}{\omega}\big)
{\bf E}_a(\omega,{\bf k})
+ \frac{1}{\omega}\big({\bf v} \cdot {\bf E}_a(\omega,{\bf k})\big) {\bf k}
\Big)\cdot \nabla_p f({\bf p})
\\[2mm] \nonumber
&+& 
\frac{g^2}{2} \int {d^3p \over (2\pi)^3} \,
\frac{{\bf v}}{\omega - {\bf v}\cdot {\bf k}}
\Big(\frac{1}{\omega} {\bf v}\times {\bf B}_{a0}({\bf k}) \Big)\cdot
\nabla_p f({\bf p})
- i g \int {d^3p \over (2\pi)^3} \,
\frac{{\bf v}}
{\omega - {\bf k}\cdot {\bf v}}\, 
\delta N^a_0({\bf k},{\bf p}) \;,
\ea
where $f({\bf p}) \equiv n({\bf p}) + \bar n ({\bf p}) 
+ 2N_c n_g({\bf p})$.

Since the chromodielectric tensor $\varepsilon^{ij}(\omega,{\bf k})$
of anisotropic plasma in the collisionless limit 
equals \cite{Mrowczynski:2000ed}
\be
\label{eij}
\varepsilon^{ij}(\omega,{\bf k}) = \delta^{ij} +
\frac{g^2}{2\omega} \int {d^3p \over (2\pi)^3} \,
\frac{v^i}{\omega - {\bf v}\cdot {\bf k}+i0^+}
\Big(\big(1-\frac{{\bf k}\cdot {\bf v}}{\omega}\big)
\delta^{jk}
+ \frac{v^jk^k}{\omega} \Big) \nabla_p^k f({\bf p})  \;,
\ee
the current can be written as 
\ba
\label{current2}
j^i_a(\omega,{\bf k}) = 
-i\omega 
\big(\varepsilon^{ij}(\omega,{\bf k}) - \delta^{ij} \big)
E^j_a(\omega,{\bf k}) 
+ \frac{g^2}{2} \int {d^3p \over (2\pi)^3} \,
\frac{{\bf v}}{\omega - {\bf v}\cdot {\bf k}}
\Big(\frac{1}{\omega} {\bf v}\times {\bf B}_{a0}({\bf k}) \Big)\cdot
\nabla_p f({\bf p})
\\[2mm]\nonumber
- i g \int {d^3p \over (2\pi)^3} \,
\frac{{\bf v}} {\omega - {\bf k}\cdot {\bf v}}\, 
\delta N^a_0({\bf k},{\bf p}) \;.
\ea
We note that the chromodielectric tensor (\ref{eij}), which
corresponds to a colorless state of the plasma, does not carry
any color indices.

Combining the third and fourth Maxwell equations (\ref{Maxwell-eqs-k}), 
one finds
\be
\label{eq3}
\big[(\omega^2 - {\bf k}^2) \, \delta^{ij}
+ k^ik^j \big] E^j(\omega,{\bf k})
= - i\omega \, j^i(\omega,{\bf k})
+i \omega E_0^i({\bf k})
-i \big({\bf k} \times {\bf B}_0({\bf k})\big)^i \;.
\ee
Substituting the current (\ref{current2}) into Eq.~(\ref{eq3}),
one obtains
\ba
\label{E-field2}
\big[ - {\bf k}^2 \delta^{ij} + k^ik^j 
+ \omega^2 \varepsilon^{ij}(\omega,{\bf k}) \big] E^j_a(\omega,{\bf k})
= - i\frac{g^2}{2} \int {d^3p \over (2\pi)^3} \,
\frac{v^i}{\omega - {\bf v}\cdot {\bf k}}
\big({\bf v}\times {\bf B}_{a0}({\bf k})\big)^j
\nabla_p^j f({\bf p})
\\[2mm] \nonumber 
- g \omega \int {d^3p \over (2\pi)^3} \,
\frac{v^i} {\omega - {\bf k}\cdot {\bf v}}\, 
\delta N^a_0({\bf k},{\bf p})
+ i \omega E_{a0}^i({\bf k})
-i \big({\bf k} \times {\bf B}_{a0}({\bf k})\big)^i \;.
\ea

Denoting the matrix in left-hand-side of Eq.~(\ref{E-field2}) as
\be
\label{matrix-sigma}
\Sigma^{ij}(\omega,{\bf k}) \equiv
- {\bf k}^2 \delta^{ij} + k^ik^j 
+ \omega^2 \varepsilon^{ij}(\omega,{\bf k}) \;,
\ee
the electric field given by Eq.~(\ref{E-field2}) can be written 
down as
\ba
\label{E-field-final}
E^i_a(\omega,{\bf k})
= - i\frac{g^2}{2} \int {d^3p \over (2\pi)^3} \,
\frac{(\Sigma^{-1})^{ij}(\omega,{\bf k}) v^j}
{\omega - {\bf v}\cdot {\bf k}}
\big({\bf v}\times {\bf B}_{a0}({\bf k})\big) 
\cdot \nabla_p f({\bf p})
- g \omega \int {d^3p \over (2\pi)^3} \,
\frac{(\Sigma^{-1})^{ij}(\omega,{\bf k}) v^j}
{\omega - {\bf k}\cdot {\bf v}}\, 
\delta N^a_0({\bf k},{\bf p})
\\[2mm] \nonumber 
+i \omega (\Sigma^{-1})^{ij}(\omega,{\bf k}) E_{a0}^j({\bf k})
-i (\Sigma^{-1})^{ij}(\omega,{\bf k})
\big({\bf k} \times {\bf B}_{a0}({\bf k})\big)^j \;.
\ea
which is the main result of this section.

When the plasma stationary state described by $f({\bf p})$
is isotropic, the dielectric tensor can be expressed through
its longitudinal and transverse components
\be
\varepsilon^{ij}(\omega,{\bf k}) = 
\varepsilon_L(\omega,{\bf k}) \: \frac{k^ik^j}{{\bf k}^2}
+ \varepsilon_T(\omega,{\bf k}) \:
\Big(\delta^{ij} - \frac{k^ik^j}{{\bf k}^2}\Big) \;,
\ee
where $\varepsilon_L(\omega,{\bf k})$ and  
$\varepsilon_T(\omega,{\bf k})$ are well known to be 
\be
\label{eL}
\varepsilon_L(\omega,{\bf k}) = 1+ \frac{g^2}{2{\bf k}^2}
\int {d^3p \over (2\pi)^3} \frac{1}
{\omega - {\bf k} \cdot {\bf v}+i0^+}
{\bf k} \cdot \frac{\partial f({\bf p})}{\partial{\bf p}} \;,
\ee
\be
\label{eT}
\varepsilon_T(\omega,{\bf k}) = 1+ \frac{g^2}{4\omega}
\int {d^3p \over (2\pi)^3} \frac{1}
{\omega - {\bf k} \cdot {\bf v}+i0^+}
\bigg[
{\bf v} \cdot \frac{\partial f({\bf p})}{\partial{\bf p}} 
- \frac{{\bf k} \cdot {\bf v}}{{\bf k}^2}
{\bf k} \cdot \frac{\partial f({\bf p})}{\partial{\bf p}} 
\bigg]\;.
\ee
The matrix $\Sigma^{ij}(\omega,{\bf k})$, which then equals
\be
\Sigma^{ij}(\omega,{\bf k}) = 
\omega^2\varepsilon_L(\omega,{\bf k})
\frac{k^ik^j}{{\bf k}^2}
+ \big(\omega^2 \varepsilon_T(\omega,{\bf k})-{\bf k}^2\big)
\Big(\delta^{ij} - \frac{k^ik^j}{{\bf k}^2}\Big) \;,
\ee
can be inverted as 
\be
\label{inv-sigma}
(\Sigma^{-1})^{ij}(\omega,{\bf k}) = 
\frac{1}{\omega^2 \varepsilon_L(\omega,{\bf k})}
\frac{k^ik^j}{{\bf k}^2}
+ \frac{1}{\omega^2 \varepsilon_T(\omega,{\bf k})-{\bf k}^2}
\Big(\delta^{ij} - \frac{k^ik^j}{{\bf k}^2}\Big) \;.
\ee

When the momentum distribution $f({\bf p})$ is isotropic, 
$\nabla_p f({\bf p}) \sim {\bf p}$, and consequently
$\big({\bf v}\times {\bf B}_0({\bf k})\big) \cdot
\nabla_p f({\bf p}) = 0$. Therefore, the first term in
the right-hand-side of Eq.~(\ref{E-field-final}) vanishes
and the electric field is found as
\ba
\label{E-field-final-iso}
E^i_a(\omega,{\bf k})
= &-& g \omega \Bigg(
\frac{1}{\omega^2 \varepsilon_L(\omega,{\bf k})}
\frac{k^ik^j}{{\bf k}^2}
+ \frac{1}{\omega^2 \varepsilon_T(\omega,{\bf k})-{\bf k}^2}
\Big(\delta^{ij} - \frac{k^ik^j}{{\bf k}^2}\Big)
\Bigg)\int {d^3p \over (2\pi)^3} \,
\frac{v^j}
{\omega - {\bf k}\cdot {\bf v}}\, 
\delta N^a_0({\bf k},{\bf p})
\\[2mm]\nonumber 
&+& i \omega \Bigg(
\frac{1}{\omega^2 \varepsilon_L(\omega,{\bf k})}
\frac{k^ik^j}{{\bf k}^2}
+ \frac{1}{\omega^2 \varepsilon_T(\omega,{\bf k})-{\bf k}^2}
\Big(\delta^{ij} - \frac{k^ik^j}{{\bf k}^2}\Big)
\Bigg) E_{a0}^j({\bf k})
- \frac{i\big({\bf k} \times {\bf B}_{a0}({\bf k})\big)^i} {\omega^2 \varepsilon_T(\omega,{\bf k})-{\bf k}^2} \;.
\ea

If the field is purely longitudinal, 
$$
{\bf E}(\omega,{\bf k}) = 
\big({\bf k}\cdot {\bf E}(\omega,{\bf k})\big) \,
\frac{\bf k}{{\bf k}^2} 
\;,\;\;\;\;
{\bf E}_0({\bf k}) = 
\big({\bf k}\cdot {\bf E}_0({\bf k})\big) \,
\frac{\bf k}{{\bf k}^2} \;,
$$
Eq.~(\ref{E-field-final-iso}) gives
\ba
\label{E-field-final-iso-long}
{\bf k}\cdot {\bf E}_a (\omega,{\bf k}) 
= - \frac{g}{\omega \varepsilon_L(\omega,{\bf k})}
\int {d^3p \over (2\pi)^3} \,
\frac{{\bf k}\cdot {\bf v}}
{\omega - {\bf k}\cdot {\bf v}}\, 
\delta N^a_0({\bf k},{\bf p})
+ \frac{i{\bf k}\cdot {\bf E}_{a0}({\bf k})}
{\omega \varepsilon_L(\omega,{\bf k})}\;.
\ea
Taking into account that 
$$ 
i{\bf k}\cdot {\bf E}_{a0}({\bf k}) 
= \rho_{a0}({\bf k})
= - g \int {d^3p \over (2\pi)^3} \,
\delta N^a_0({\bf k},{\bf p}) \;,
$$
Eq.~(\ref{E-field-final-iso-long}) can be rewritten as
\ba
\label{E-field-final-iso-long2}
{\bf k}\cdot {\bf E}_a(\omega,{\bf k}) 
= - \frac{g}{\varepsilon_L(\omega,{\bf k})}
\int {d^3p \over (2\pi)^3} \,
\frac{1}{\omega - {\bf k}\cdot {\bf v}} \:
\delta N^a_0({\bf k},{\bf p}) \;.
\ea
Eq.~(\ref{E-field-final-iso-long2}) can be obtained directly 
by substituting the solution of transport equation 
(\ref{solution1}) (with ${\bf B}=0$) into the first 
Maxwell equation. Then, the initial electric field
does not show up.

\subsection{Chromomagnetic field}

Using again the third Maxwell equation (\ref{Maxwell-eqs-k}) to 
express the magnetic field through the electric one, 
Eq.~(\ref{E-field-final}) immediately provides
\ba
\label{B-field-final}
B^i_a(\omega,{\bf k}) = \frac{1}{\omega}
\epsilon^{ijk}k^j
(\Sigma^{-1})^{kl}(\omega,{\bf k})
\Bigg(- i \frac{g^2}{2} \int {d^3p \over (2\pi)^3} \,
\frac{v^l}{\omega - {\bf v}\cdot {\bf k}}
\big({\bf v}\times {\bf B}_{a0}({\bf k})\big)
\cdot \nabla_p f({\bf p})
\\[2mm] \nonumber 
- g \omega \int {d^3p \over (2\pi)^3} \,
\frac{v^l}
{\omega - {\bf k}\cdot {\bf v}}\, 
\delta N^a_0({\bf k},{\bf p})
+i \omega E_{a0}^l({\bf k})
-i \big({\bf k} \times {\bf B}_{a0}({\bf k})\big)^l
\Bigg) + \frac{i}{\omega}B_{a0}^i({\bf k})
\;.
\ea
When the plasma stationary state is isotropic and 
$(\Sigma^{-1})^{ij}(\omega,{\bf k})$ is given by Eq.~(\ref{inv-sigma}),
one finds
\be
\epsilon^{ijk}k^j
(\Sigma^{-1})^{kl}(\omega,{\bf k}) = 
\frac{\epsilon^{ijl}k^j}
{\omega^2 \varepsilon_T(\omega,{\bf k})-{\bf k}^2}
\;.
\ee
The first term in the right-hand-side of Eq.~(\ref{B-field-final})
vanishes, because $\big({\bf v}\times {\bf B}_0({\bf k})\big) \cdot
\nabla_p f({\bf p}) =0$, and thus
\ba
\label{B-field-final-iso}
{\bf B}_a(\omega,{\bf k}) = 
- \frac{g}
{\big(\omega^2 \varepsilon_T(\omega,{\bf k})-{\bf k}^2\big)}
\int {d^3p \over (2\pi)^3} \,
\frac{{\bf k} \times {\bf v}}{\omega - {\bf k}\cdot {\bf v}}\,
\delta N^a_0({\bf k},{\bf p}) 
&+& \frac{i{\bf k} \times {\bf E}_{a0}({\bf k})}
{\omega^2 \varepsilon_T(\omega,{\bf k})-{\bf k}^2}
\\[2mm]\nonumber
&+& \frac{i\omega \varepsilon_T(\omega,{\bf k})}
{\omega^2 \varepsilon_T(\omega,{\bf k})-{\bf k}^2}
{\bf B}_{a0}({\bf k})
\;.
\ea


\section{Initial Fluctuations}
\label{sec-initial-fluc}


The correlation functions 
$\langle E^i_a(t_1,{\bf r}_1) E^j_b(t_2,{\bf r}_2) \rangle$,
$\langle B^i_a(t_1,{\bf r}_1) B^j_b(t_2,{\bf r}_2) \rangle$,
where $\langle \cdots \rangle$ denotes averaging over statistical
ensemble, are determined by the initial correlations such as
$\langle \delta N_0^a({\bf r}_1,{\bf p}_1) 
\delta N_0^b({\bf r}_2,{\bf p}_2)\rangle$, 
$\langle E_{a0}^i({\bf r}_1) E_{b0}^j({\bf r}_2) \rangle$,
$\langle \delta N_0^a({\bf r}_1,{\bf p}_1) 
E_{b0}^j({\bf r}_2) \rangle$ {\it etc.},
which are discussed in this section. 

The initial correlation of the distribution functions 
$\langle \delta Q_0^{mn}({\bf r}_1,{\bf p}_1) 
\delta Q_0^{pr}({\bf r}_1,{\bf p}_1)\rangle$
is assumed to be given by the correlation function 
$\langle \delta Q^{mn}(t_1,{\bf r}_1,{\bf p}_1) 
\delta Q^{pr}(t_2,{\bf r}_2,{\bf p}_2)\rangle_{\rm free}$
taken at $t_1 = t_2 = 0$ of the classical system of free quarks 
in a stationary homogeneous state described by the distribution 
function $Q^0({\bf p})$. Such a correlation function of particles, 
which obey Boltzmann statistics and have no internal degrees of 
freedom, is well known to be \cite{LP81}
\be
\label{corr-free}
\langle \delta f(t_1,{\bf r}_1,{\bf p}_1) 
\delta f(t_2,{\bf r}_2,{\bf p}_2)\rangle_{\rm free}
= (2\pi )^3 \delta^{(3)}({\bf p}_1 - {\bf p}_2) \,
\delta^{(3)}\big({\bf r}_2 - {\bf r}_1 
- {\bf v}_1(t_2 - t_1)\big) \: f^0({\bf p}_1) \;,
\ee
where $ f^0({\bf p}) \equiv 
\langle f(t,{\bf r},{\bf p}) \rangle_{\rm free}$. 
The correlation expressed by Eq.~(\ref{corr-free}) occurs when 
the same particle travels from the space-time point $(t_1,{\bf r}_1)$ 
to $(t_2, {\bf r}_2)$. 

A generalization of the formula (\ref{corr-free}) to the case
of quarks and gluons carrying classical color charges was discussed
in \cite{Litim:1999id,Litim:2001db}. In the Appendix we give a 
quantum mechanical and relativistic derivation of the correlation 
function of the distribution functions of free quarks and gluons 
with the matrix color degrees of freedom. The results is valid
for equilibrium and nonequilibrium systems. In the classical limit 
the correlation functions equal
\ba
\label{dQ-dQ}
\langle \delta Q^{mn}(t_1,{\bf r}_1,{\bf p}_1) 
\delta Q^{pr}(t_2,{\bf r}_2,{\bf p}_2)\rangle_{\rm free}
= \delta^{mr} \delta^{np} 
(2\pi )^3 \delta^{(3)}({\bf p}_1 - {\bf p}_2) \,
\delta^{(3)}\big({\bf r}_2 - {\bf r}_1 
- {\bf v}_1(t_2 - t_1)\big) \: n({\bf p}_1) \;,
\\[2mm] 
\label{dbarQ-dbarQ}
\langle \delta \bar Q^{mn}(t_1,{\bf r}_1,{\bf p}_1) 
\delta \bar Q^{pr}(t_2,{\bf r}_2,{\bf p}_2)\rangle_{\rm free} 
= \delta^{mr} \delta^{np} 
(2\pi )^3 \delta^{(3)}({\bf p}_1 - {\bf p}_2) \,
\delta^{(3)}\big({\bf r}_2 - {\bf r}_1 
- {\bf v}_1(t_2 - t_1)\big) \: \bar n({\bf p}_1) \;,
\\[2mm] 
\label{dG-dG}
\langle \delta G^{ab}(t_1,{\bf r}_1,{\bf p}_1) 
\delta G^{cd}(t_2,{\bf r}_2,{\bf p}_2)\rangle_{\rm free}
= \delta^{ad} \delta^{bc} 
(2\pi )^3 \delta^{(3)}({\bf p}_1 - {\bf p}_2) \,
\delta^{(3)}\big({\bf r}_2 - {\bf r}_1 
- {\bf v}_1(t_2 - t_1)\big) \: n_g({\bf p}_1) \;,
\ea
where, as previously, the color indices $m,n,p,r$ refer to
the fundamental representation while the indices $a,b,c,d$ 
to the adjoint one. The correlation functions of the 
distribution functions of different particles such as 
$\langle \delta Q(t_1,{\bf r}_1,{\bf p}_1) 
\delta G(t_2,{\bf r}_2,{\bf p}_2)\rangle_{\rm free}$ vanish.

The initial correlation of the function 
$\delta N^a(t,{\bf r},{\bf p})$ defined by 
Eq.~(\ref{N-def}) is provided by 
Eqs.~(\ref{dQ-dQ}, \ref{dbarQ-dbarQ}, \ref{dG-dG}) as
\ba
\langle \delta N_0^a({\bf r}_1,{\bf p}_1) 
\delta N_0^b({\bf r}_1,{\bf p}_1)\rangle
&=& \langle \delta N^a (t_1=0,{\bf r}_1,{\bf p}_1) 
\delta N^b(t_2=0,{\bf r}_2,{\bf p}_2)\rangle_{\rm free}
\\[2mm] \nonumber
&=& \frac{1}{2} \delta^{ab}
(2\pi )^3 \delta^{(3)}({\bf p}_1 - {\bf p}_2) \,
\delta^{(3)}({\bf r}_1 - {\bf r}_2) \: f({\bf p}_1) \;,
\ea
where, as previously, $f({\bf p}) \equiv n({\bf p}) 
+ \bar n({\bf p}) + 2N_c n_g({\bf p})$. The Fourier
transform with respect to the space variables equals
\be
\label{NN-0}
\langle \delta N_0^a ({\bf k}_1,{\bf p}_1) 
\delta N_0^b ({\bf k}_2,{\bf p}_2) \rangle
= \frac{1}{2} \delta^{ab}
(2\pi )^3\delta^{(3)}({\bf p}_1 - {\bf p}_2) \:
(2\pi )^3 \delta^{(3)}({\bf k}_1 + {\bf k}_2) \: f({\bf p}_1) \;.
\ee

To compute the correlations functions like
$\langle E_{a0}^i({\bf r}_1) E_{b0}^j({\bf r}_2) \rangle$,
$\langle \delta Q_0^{mn}({\bf r}_1,{\bf p}_1) 
E_{a0}^j({\bf r}_2) \rangle$ or 
$\langle E_{a0}^i({\bf r}_1) B_{b0}^j({\bf r}_2) \rangle$,
we use the Maxwell equations transformed using the
Fourier transformation not the one-sided Fourier 
transformation. Actually, the Fourier transformed Maxwell
equations are very similar to the one-sided Fourier 
transformed Maxwell equations (\ref{Maxwell-eqs-k}).
The initial electric and magnetic fields are simply
absent in the former ones. However, it should be clearly 
stated that the one-sided Fourier transformation is 
{\em not} mixed up with the Fourier transformation. The 
latter is used to compute only the initial fluctuations 
which are independent of $\omega$.

Combining the third and the fourth Maxwell equation, one 
gets the equation as Eq.~(\ref{eq3}) but the terms
with ${\bf E}_0({\bf k})$ and ${\bf B}_0({\bf k})$ are
absent. Inverting the matrix in the right-hand-side of 
the equation, we get the electric field expressed through
the current
\be
E^i(\omega,{\bf k})
= - i\omega 
\bigg[ \frac{1}{\omega^2} 
\frac{k^ik^j}{{\bf k}^2} + 
\frac{1}{\omega^2 - {\bf k}^2} \, \Big(
\delta^{ij} - \frac{k^ik^j}{{\bf k}^2}\Big)
\bigg] \, j^j(\omega,{\bf k}) \;.
\ee
The magnetic field is given as
\be
\label{B-field-2}
{\bf B}(\omega,{\bf k})
= - \frac{i}{\omega^2 - {\bf k}^2} \;  
{\bf k} \times {\bf j}(\omega,{\bf k})\;.
\ee

The correlation function 
$\langle E_{a0}^i({\bf k}_1) E_{b0}^j({\bf k}_2) \rangle$
is derived as  
\ba
\label{eq-EE-0-77}
\langle E_{a0}^i({\bf k}_1) E_{b0}^j({\bf k}_2) \rangle
&=& \int \frac{d\omega_1}{2\pi}\frac{d\omega_2}{2\pi}
\langle E_a^i(\omega_1,{\bf k}_1) E_b^j(\omega_2,{\bf k}_2) \rangle
= - \int \frac{d\omega_1}{2\pi}\frac{d\omega_2}{2\pi}
\\[2mm] \nonumber
&\times&  \bigg[ \frac{1}{\omega_1} 
\frac{k^i_1k^k_1}{{\bf k}^2_1} + 
\frac{\omega_1}{\omega^2_1 - {\bf k}^2_1} \, 
\Big(\delta^{ik} - \frac{k^i_1k^k_1}{{\bf k}^2_1}\Big)
\bigg]
\bigg[\frac{1}{\omega_2} 
\frac{k^j_2k^l_2}{{\bf k}^2_2} + 
\frac{\omega_2}{\omega^2_2 - {\bf k}^2_2} \, 
\Big(\delta^{jl} - \frac{k^j_1k^l_1}{{\bf k}^2_1}\Big)
\bigg]
\\[2mm] \nonumber
&\times& \langle j_a^k(\omega_1,{\bf k}_1) j_b^j(\omega_2,{\bf k}_2) \rangle
\\[2mm] \nonumber
&=& - g^2 \int \frac{d\omega_1}{2\pi}\frac{d\omega_2}{2\pi}
\frac{d^3p_1}{(2\pi)^3}\frac{d^3p_2}{(2\pi)^3}\, v^k_1 v^l_2 
\\[2mm] \nonumber
&\times& 
\bigg[ \frac{1}{\omega_1} 
\frac{k^i_1k^k_1}{{\bf k}^2_1} + 
\frac{\omega_1}{\omega^2_1 - {\bf k}^2_1} \, 
\Big(\delta^{ik} - \frac{k^i_1k^k_1}{{\bf k}^2_1}\Big)
\bigg]
\bigg[ \frac{1}{\omega_2} 
\frac{k^j_2k^l_2}{{\bf k}^2_2} + 
\frac{\omega_2}{\omega^2_2 - {\bf k}^2_2} \, 
\Big(\delta^{jl} - \frac{k^j_1k^l_1}{{\bf k}^2_1}\Big)
\bigg]
\\[2mm] \nonumber
&\times& 
\langle \delta N^a(\omega_1,{\bf k}_1,{\bf p}_1) 
\delta N^b(\omega_2,{\bf k}_2,{\bf p}_2) \rangle  \;.
\ea

Using the formulas
(\ref{dQ-dQ}, \ref{dbarQ-dbarQ}, \ref{dG-dG}), one easily 
finds the Fourier transform of the correlation function
of $\delta N^a$ as
\ba
\label{N-N}
\langle \delta N^a(\omega_1,{\bf k}_1,{\bf p}_1) 
\delta N^b(\omega_2,{\bf k}_2,{\bf p}_2) \rangle_{\rm free}
= \frac{1}{2}\delta^{ab}  
 (2\pi )^3\delta^{(3)}({\bf p}_2 - {\bf p}_1) \:
2\pi \delta (\omega_1 + \omega_2) \:
(2\pi )^3 \delta^{(3)}({\bf k}_1 + {\bf k}_2) \: 
\\[2mm]\nonumber
\times 
2\pi\delta
\Big(\frac{\omega_1 - \omega_2}{2} 
- \frac{{\bf k}_1-{\bf k}_2}{2}{\bf v}_1\Big) \: 
f({\bf p}_1) \;.
\ea

Substituting the formula (\ref{N-N}) into Eq.~(\ref{eq-EE-0-77})
and performing trivial integrations, one finally obtains
\ba
\label{EE-0}
\langle E_{a0}^i({\bf k}_1) E_{b0}^j({\bf k}_2) \rangle
= -\frac{g^2}{2} \: \delta^{ab} 
(2\pi )^3 \delta^{(3)}({\bf k}_1 + {\bf k}_2) 
\int \frac{d^3p}{(2\pi)^3} \: f({\bf p}) \:
\frac{\big(({\bf k}_1 \cdot {\bf v})v^i - k^i_1\big)
\big(({\bf k}_2 \cdot {\bf v})v^j - k^j_2\big)}
{\big(({\bf k}_1 \cdot {\bf v})^2 - {\bf k}_1^2\big)
\big(({\bf k}_2 \cdot {\bf v})^2 - {\bf k}_2^2\big)}
\;.
\ea

Analogously to the correlation function 
$\langle E_{a0}^i({\bf k}_1) E_{b0}^j({\bf k}_2) \rangle$, 
one finds
\be
\langle E_{a0}^i({\bf k}_1) 
\delta N_{b0}({\bf k}_2,{\bf p}_2) \rangle
= i \frac{g}{2} \: \delta^{ab}
(2\pi )^3 \delta^{(3)}({\bf k}_1 + {\bf k}_2) \: 
f({\bf p}_2) \:
\frac{({\bf k}_1 \cdot {\bf v}_2)v^i_2 - k^i_1}
{({\bf k}_1 \cdot {\bf v}_2)^2 - {\bf k}_1^2} \;.
\ee

Starting with Eq.~(\ref{B-field-2}), we obtain
\be
\label{BB-0}
\langle B_{a0}^i({\bf k}_1) B_{b0}^j({\bf k}_2) \rangle
= - \frac{g^2}{2} \: \delta^{ab}
(2\pi )^3 \delta^{(3)}({\bf k}_1 + {\bf k}_2) \:
\epsilon^{ikl} \epsilon^{jmn} k^k_1 k^m_2
\int \frac{d^3p}{(2\pi)^3} \: f({\bf p}) \:
\frac{v^l v^n}
{\big(({\bf k}_1 \cdot {\bf v})^2 - {\bf k}_1^2\big)
\big(({\bf k}_2 \cdot {\bf v})^2 - {\bf k}_2^2\big)}
\;,
\ee
and 
\be
\langle B_{a0}^i({\bf k}_1) 
\delta N_0^b({\bf k}_2,{\bf p}_2) \rangle
= i \frac{g}{2} \: \delta^{ab}
(2\pi )^3 \delta^{(3)}({\bf k}_1 + {\bf k}_2) \: 
f({\bf p}_2) \:
\frac{\epsilon^{ijk}k^j_1 v^k_2}
{({\bf k}_1 \cdot {\bf v}_2)^2 - {\bf k}_1^2} \;.
\ee
Finally, one computes
\be
\label{EB-0}
\langle E_{a0}^i({\bf k}_1) B_{b0}^j({\bf k}_2) \rangle
= - \frac{g^2}{2} \: \delta^{ab}
(2\pi )^3 \delta^{(3)}({\bf k}_1 + {\bf k}_2) 
\int \frac{d^3p}{(2\pi)^3} \: f({\bf p}) \:
\frac{\big(({\bf k}_1 \cdot {\bf v})v^i - k^i_1\big)
\epsilon^{jkl} k^k_2 v^l}
{\big(({\bf k}_1 \cdot {\bf v})^2 - {\bf k}_1^2\big)
\big(({\bf k}_2 \cdot {\bf v})^2 - {\bf k}_2^2\big)}
\;.
\ee


\section{Chromomagnetic Field in Isotropic QGP}
\label{sec-fluc-B}


As seen in Eq.~(\ref{B-field-final-iso}), the magnetic field in 
isotropic plasma is given by three terms. Therefore, nine terms 
enter the correlation function 
$\langle B_a^i(\omega_1,{\bf k}_1) B_b^j(\omega_2,{\bf k}_2)\rangle$.
Substituting into these terms the initial fluctuations derived in
Sec.~\ref{sec-initial-fluc}, one finds after an elementary but 
lengthy and tedious analysis the following expression
\ba
\label{B-fluc-iso3}
\langle B_a^i(\omega_1,{\bf k}_1) B_b^j(\omega_2,{\bf k}_2)\rangle 
&=& 
\frac{g^2}{2}\: \delta^{ab}
\frac{(2\pi )^3\delta^{(3)}({\bf k}_1 + {\bf k}_2)
\: \epsilon^{ikl}\epsilon^{jmn} k_1^k k_2^m }
{\big(\omega_1^2 \varepsilon_T(\omega_1,{\bf k}_1)-{\bf k}_1^2)\big)
\big(\omega_2^2 \varepsilon_T(\omega_2,{\bf k}_2)-{\bf k}_2^2)\big)}
\\[2mm] \nonumber
&\times& 
\int {d^3p \over (2\pi)^3}\, f({\bf p})
\frac{v^l v^n}{
(\omega_1 - {\bf k}_1\cdot {\bf v})
(\omega_2 - {\bf k}_2\cdot {\bf v})
(({\bf k}_1\cdot {\bf v})^2 -{\bf k}_1^2)
(({\bf k}_2\cdot {\bf v})^2 -{\bf k}_2^2)}
\\[2mm] \nonumber
&\times& 
\Big[
(\omega_1 ({\bf k}_1\cdot {\bf v}) - {\bf k}_1^2)
+ \omega_1 \varepsilon_T(\omega_1,{\bf k}_1)
(\omega_1 - {\bf k}_1\cdot {\bf v}) 
\Big]
\\[2mm] \nonumber
&\times&
\Big[
(\omega_2 ({\bf k}_2\cdot {\bf v}) - {\bf k}_2^2)
+ \omega_2 \varepsilon_T(\omega_2,{\bf k}_2)
(\omega_2 - {\bf k}_2\cdot {\bf v}) 
\Big] \;.
\ea

We now compute 
$\langle B_a^i(t_1,{\bf r}_1)B_b^j(t_2,{\bf r}_2) \rangle$ given by 
\ba
\langle B_a^i(t_1,{\bf r}_1) B_b^j(t_2,{\bf r}_2) \rangle
&=& \int_{-\infty +i\sigma}^{\infty +i\sigma}
{d\omega_1 \over 2\pi}
\int_{-\infty +i\sigma}^{\infty +i\sigma}
{d\omega_2 \over 2\pi}
\int {d^3k_1 \over (2\pi)^3}
\int {d^3k_2 \over (2\pi)^3}
\\[2mm] \nonumber
&\times& e^{-i(\omega_1 t_1 - {\bf k}_1\cdot {\bf r}_1
+ \omega_2 t_2 - {\bf k}_2\cdot {\bf r}_2)}
\langle B_a^i(\omega_1,{\bf k}_1)
B_b^j(\omega_2,{\bf k}_2) \rangle \;.
\ea
Zeros of $(\omega_i^2\varepsilon_T(\omega_i,{\bf k}_i)- {\bf k}_i^2)$ 
and of $\omega_i - {\bf k}_i \cdot {\bf v} + i0^+)$ 
with $i=1,2$ contribute to the integrals over $\omega_1$ and 
$\omega_2$. The equation 
$\omega^2\varepsilon_T(\omega,{\bf k}) - {\bf k}^2 = 0$
determines the plasma collective transverse modes, while
$\omega - {\bf k} \cdot {\bf v} = 0$ corresponds to the interaction 
of plasma particles of velocity ${\bf v}$ with the modes of phase
velocity ${\bf v}_\phi \equiv \omega /|{\bf k}|$. Since the plasma 
system under consideration is stable with respect to transverse 
modes - the modes are expected to be damped, all zeros 
of $(\omega_i^2\varepsilon_T(\omega_i,{\bf k}_i)- {\bf k}_i^2)$ 
lie in the lower half-plane of complex $\omega$. Consequently, the 
contributions associated with these zeros exponentially decay in 
time and they vanish in the long-time limit of both $t_1$ and $t_2$.
The long-time limit corresponds to times which are much longer
than the decay time of collective excitations in the plasma
\footnote{The collective modes, which are obtained with the dielectric 
functions (\ref{eL}, \ref{eT}), are actually not damped, see e.g. 
\cite{Thoma:1995ju}. The damping appears to be a higher order effect.}.

We further consider the long-time limit of 
$\langle B_a^i(t_1,{\bf r}_1) B_b^j(t_2,{\bf r}_2) \rangle$ and 
then, the only non-vanishing contribution is related to the 
poles at $\omega_1 = {\bf k}_1 \cdot {\bf v}$ and 
$\omega_2 = {\bf k}_2 \cdot {\bf v}$. This contribution reads
\ba
\langle B_a^i(t_1,{\bf r}_1) B_b^j(t_2,{\bf r}_2) \rangle_\infty
&=& - \frac{g^2}{2} \: \delta^{ab} \int 
{d^3k_1 \over (2\pi)^3}
{d^3k_2 \over (2\pi)^3}
{d^3p \over (2\pi)^3}\, f({\bf p}) \;
e^{-i(\omega_1 t_1 - {\bf k}_1\cdot {\bf r}_1
+ \omega_2 t_2 - {\bf k}_2\cdot {\bf r}_2)}
\\[2mm] \nonumber
&\times& 
\frac{(2\pi )^3\delta^{(3)}({\bf k}_1 + {\bf k}_2)
\: \epsilon^{ikl}\epsilon^{jmn} k_1^k k_2^m }
{\big(\omega_1^2 \varepsilon_T(\omega_1,{\bf k}_1)-{\bf k}_1^2)\big)
\big(\omega_2^2 \varepsilon_T(\omega_2,{\bf k}_2)-{\bf k}_2^2)\big)}
\\[2mm] \nonumber
&\times& 
\frac{v^l v^n}{
(({\bf k}_1\cdot {\bf v})^2 -{\bf k}_1^2)
(({\bf k}_2\cdot {\bf v})^2 -{\bf k}_2^2)}
(\omega_1 ({\bf k}_1\cdot {\bf v}) - {\bf k}_1^2) (\omega_2 ({\bf k}_2\cdot {\bf v}) - {\bf k}_2^2)
\Bigg|_{\omega_1 = {\bf k}_1 \cdot {\bf v},\;\;\;
\omega_2 = {\bf k}_2 \cdot {\bf v}}
\;.
\ea
It can be easily expressed as 
\ba
\langle B_a^i(t_1,{\bf r}_1) B_b^j(t_2,{\bf r}_2) \rangle_\infty 
&=& 
\int {d\omega \over 2\pi} {d^3k \over (2\pi)^3}
e^{-i \big(\omega (t_1 - t_2)
 - {\bf k}\cdot ({\bf r}_1 - {\bf r}_2)\big)}
\langle B_a^i B_b^j\rangle_{\omega \, {\bf k}} \;,
\ea
where the fluctuation spectrum is
\ba
\langle B_a^i B_b^j\rangle_{\omega \, {\bf k}}
= 
\frac{\pi \, g^2 \delta^{ab} \epsilon^{ikl}\epsilon^{jmn} k^k k^m}
{\big(\omega^2 \varepsilon_T(\omega,{\bf k})-{\bf k}^2)\big)
\big(\omega^2 \varepsilon_T(-\omega,-{\bf k})-{\bf k}^2)\big)}
\int {d^3p \over (2\pi)^3} \, f({\bf p}) \:
\delta (\omega - {\bf k} \cdot {\bf v}) \: v^l v^n \;.
\ea
When both $\omega$ and ${\bf k}$ are real 
$\varepsilon_T(-\omega,-{\bf k}) = \varepsilon_T^*(\omega,{\bf k})$.
Therefore, the fluctuation spectrum can be rewritten as
\ba
\label{BiBj-spec}
\langle B_a^i B_b^j\rangle_{\omega \, {\bf k}}
=\frac{\pi g^2 \delta^{ab} \epsilon^{ikl}\epsilon^{jmn} k^k k^m}
{\big|\omega^2 \varepsilon_T(\omega,{\bf k})-{\bf k}^2\big|^2}
\int {d^3p \over (2\pi)^3}\, f ({\bf p}) \:
\delta (\omega - {\bf k} \cdot {\bf v}) \: v^l v^n
\;.
\ea

One observes that the matrix function 
\be
\label{M-def}
M^{ij}(\omega,{\bf k}) \equiv
\int {d^3p \over (2\pi)^3}\, f ({\bf p}) \:
\delta (\omega - {\bf k} \cdot {\bf v}) \: v^i v^j \;,
\ee
which enters the correlation function (\ref{BiBj-spec}), can 
be decomposed as
\be
\label{M-L-T}
M^{ij}(\omega,{\bf k}) = 
M_L(\omega,{\bf k}) \: \frac{k^ik^j}{{\bf k}^2} + 
M_T(\omega,{\bf k}) \:
\Big(\delta^{ij} - \frac{k^ik^j}{{\bf k}^2}\Big) \;,
\ee
because the plasma is assumed to be isotropic. Comparing 
Eq.~(\ref{M-L-T}) to Eq.~(\ref{M-def}), one finds
\be
\label{M_L}
M_L(\omega,{\bf k}) \equiv
\int {d^3p \over (2\pi)^3}\, f({\bf p}) \:
\delta (\omega - {\bf k} \cdot {\bf v}) \: 
\frac{({\bf k}\cdot {\bf v})^2}{{\bf k}^2}
\;,
\ee
\be
\label{M_T}
M_T(\omega,{\bf k}) \equiv
\frac{1}{2}
\int {d^3p \over (2\pi)^3}\, f({\bf p}) \:
\delta (\omega - {\bf k} \cdot {\bf v}) \, 
\bigg[{\bf v}^2 -
\frac{({\bf k}\cdot {\bf v})^2}{{\bf k}^2}
\bigg] \;.
\ee
Using the decomposition (\ref{M-L-T}), the correlation function 
(\ref{BiBj-spec}) can be written down as
\ba
\label{BiBj-M-spec}
\langle B_a^i B_b^j\rangle_{\omega \, {\bf k}}
=\frac{\pi g^2 \delta^{ab} (\delta^{ij}{\bf k}^2 - k^i k^j )}
{\big|\omega^2 \varepsilon_T(\omega,{\bf k})-{\bf k}^2\big|^2}
M_T(\omega,{\bf k})\;.
\ea

For equilibrium plasma the correlation function 
$\langle B_a^i B_b^j\rangle_{\omega \, {\bf k}}$ can be expressed 
in the form of fluctuation-dissipation relation. One first 
observes that due to the identity
\be
\label{iden1}
\frac{1}{x\pm i0^+} = {\cal P}\frac{1}{x} \mp i\pi \delta (x) \;,
\ee
the imaginary part of $\varepsilon_T(\omega,{\bf k})$, which is 
given by Eq.~(\ref{eT}), is 
\be
\Im \varepsilon_T(\omega,{\bf k}) =
- \frac{\pi g^2}{4 \omega} 
\int {d^3p \over (2\pi)^3} \:
\delta(\omega - {\bf k}\cdot {\bf v}) \:
\bigg[
{\bf v} \cdot \frac{\partial f({\bf p})}{\partial{\bf p}} 
- \frac{{\bf k} \cdot {\bf v}}{{\bf k}^2} \;
{\bf k} \cdot \frac{\partial f({\bf p})}{\partial{\bf p}} 
\bigg]\;. 
\ee
In equilibrium $f({\bf p}) \sim e^{-\beta E_p}$ and 
$\partial f({\bf p})/\partial{\bf p} = - \beta {\bf v} f({\bf p})$.
Therefore, $\Im \varepsilon_T$ equals
\be
\label{Im-eT-eq}
\Im \varepsilon_T(\omega,{\bf k}) =
\frac{\pi g^2}{4T \omega \, {\bf k}^2} \int {d^3p \over (2\pi)^3} \:
\delta(\omega - {\bf k}\cdot {\bf v}) \:
\big({\bf k}^2{\bf v}^2 
- ({\bf k} \cdot {\bf v})^2\big)\: f({\bf p})
\;,
\ee
where $T \equiv 1/\beta$ is the system's temperature.
Consequently, the function $M_T$ (\ref{M_T}) can be expressed 
through $\Im \varepsilon_T$ (\ref{Im-eT-eq}) as
\be
M_T(\omega,{\bf k}) = \frac{2 T \omega}{\pi g^2} \:
\Im \varepsilon_T(\omega,{\bf k}) \;,
\ee 
and finally,
\ba
\label{BiBj-spec-eq}
\langle B_a^i B_b^j\rangle_{\omega \, {\bf k}}
=\frac{2 T}{\omega^3} \:\delta^{ab}
(\delta^{ij}{\bf k}^2 - k^i k^j ) \:
\frac{\Im \varepsilon_T (\omega,{\bf k})}
{\big|\varepsilon_T(\omega,{\bf k})-
\frac{{\bf k}^2}{\omega^2}\big|^2} \;.
\ea
As seen, the fluctuation spectrum has strong peaks corresponding
to collective transverse modes determined by the dispersion
equation $\omega^2\varepsilon_T(\omega,{\bf k})- {\bf k}^2 =0$.
The electromagnetic counterpart of Eq.~(\ref{BiBj-spec-eq}),
which is derived in \cite{Mrow2007}, coincides with the formula 
(11.2.2.7) from \cite{Akh75} obtained there directly from the  fluctuation-dissipation theorem. When Eq.~(\ref{BiBj-spec-eq})
is compared to the electromagnetic formula one should
remember that the Gauss units are used in \cite{Akh75,Mrow2007}
while the units, which are usually applied in QCD considerations,
correspond to the Heaviside-Lorentz electromagnetic system. 
The magnetic field in the Gauss units ${\bf B}_{\rm Gauss}$ is  related to the magnetic field in the Heaviside-Lorentz units 
${\bf B}_{\rm HL}$ as 
${\bf B}_{\rm Gauss}=\sqrt{4\pi}\, {\bf B}_{\rm HL}$. We also  mention that the correlation functions summed over colors
such as $\langle B_a^i B_a^j\rangle$ are gauge independent
as shown in Sec.~\ref{subsec-gauge}. Finally, we note that 
Eq.~(\ref{BiBj-spec-eq}) remains unchanged when the effect of 
quantum statistics of quarks and gluons is incorporated. However, 
the equilibrium expression of $\Im \varepsilon_T$, which is given 
by Eq.~(\ref{Im-eT-eq}), needs to be modified as explained in 
Sec.~\ref{subsec-statistics}.

                                                                                
\section{Chromoelectric Field in Isotropic QGP}
\label{sec-fluc-E}
                                                                                

The analysis of chromoelectric field fluctuations is much more complicated 
than that of the magnetic field. First of all, there are five terms 
which enter the formula of electric field given by 
Eq.~(\ref{E-field-final-iso}), and consequently, the correlation function 
$\langle E_a^i(\omega_1,{\bf k}_1) E_b^j(\omega_2,{\bf k}_2) \rangle$ 
includes 25 terms. The magnetic field is purely transverse 
and some terms automatically drop out but the electric fields have 
longitudinal and transverse components. Using the formulas of 
initial fluctuations, which are derived in Sec.~\ref{sec-initial-fluc},  
and patiently analyzing term by term, one obtains after an elementary 
but very lengthy calculation the correlation function of the form 
\ba
\label{EiEj-1}
\langle E_a^i(\omega_1,{\bf k}_1)  E_a^j(\omega_2,{\bf k}_2) \rangle
&=& \frac{g^2}{2}\,\delta^{ab} 
(2\pi )^3\delta^{(3)}({\bf k}_1 + {\bf k}_2)
\int {d^3p \over (2\pi)^3} \, f({\bf p})
\\[2mm]\nonumber
&\times \Bigg\{&
\frac{k_1^i}{\omega_1^2 \varepsilon_L(\omega_1,{\bf k}_1)} \:
\frac{k_2^j}{\omega_2^2 \varepsilon_L(\omega_2,{\bf k}_2)} \:
\frac{\omega_1^2 \omega_2^2}
{{\bf k}_1^2 
(\omega_1 - {\bf k}_1\cdot {\bf v}) \:
{\bf k}_2^2 
(\omega_2 - {\bf k}_2\cdot {\bf v})}
\\[2mm]\nonumber
&+&
\frac{k_1^i}{\omega_1^2 \varepsilon_L(\omega_1,{\bf k}_1)} \:
\frac{ v^j {\bf k}_2^2 - k_2^j ({\bf k}_2 \cdot {\bf v})}
{\omega_2^2 \varepsilon_T(\omega_2,{\bf k}_2)-{\bf k}_2^2}\:
\frac{
\omega_1^2 
[\omega_2
(\omega_2({\bf k}_2\cdot {\bf v}) - {\bf k}_2^2)
-
{\bf k}_2^2 
(\omega_2 - {\bf k}_2\cdot {\bf v})]}
{{\bf k}_1^2
(\omega_1 - {\bf k}_1\cdot {\bf v})\:
{\bf k}_2^2
(\omega_2 - {\bf k}_2\cdot {\bf v})\:
(({\bf k}_2\cdot {\bf v})^2 - {\bf k}_2^2) }
\\[2mm]\nonumber
&+&
\frac{ v^i {\bf k}_1^2 - k_1^i ({\bf k}_1 \cdot {\bf v})}
{\omega_1^2 \varepsilon_T(\omega_1,{\bf k}_1)-{\bf k}_1^2}\:
\frac{k_2^j}{\omega_2^2 \varepsilon_L(\omega_2,{\bf k}_2)} \:
\frac{
\omega_2^2 
[\omega_1 
(\omega_1 ({\bf k}_1\cdot {\bf v}) - {\bf k}_1^2)
-
{\bf k}_1^2
(\omega_1 - {\bf k}_1\cdot {\bf v})]}
{{\bf k}_1^2
(\omega_1 - {\bf k}_1\cdot {\bf v})\:
(({\bf k}_1\cdot {\bf v})^2 - {\bf k}_1^2)\:
{\bf k}_2^2
(\omega_2 - {\bf k}_2\cdot {\bf v})}
\\[2mm]\nonumber
&+&
\frac{k_1^i ({\bf k}_1\cdot {\bf v}) - v^i{\bf k}_1^2}
{\omega_1^2 \varepsilon_T(\omega_1,{\bf k}_1)-{\bf k}_1^2} \:
\frac{k_2^j ({\bf k}_2\cdot {\bf v}) - v^j{\bf k}_2^2}
{\omega_2^2 \varepsilon_T(\omega_2,{\bf k}_2)-{\bf k}_2^2}
\\[2mm]\nonumber
&&\times
\frac{
\omega_1 (\omega_1 ({\bf k}_1\cdot {\bf v}) - {\bf k}_1^2)
-
{\bf k}_1^2 (\omega_1 - {\bf k}_1\cdot {\bf v})}
{{\bf k}_1^2
(\omega_1 - {\bf k}_1\cdot {\bf v})\:
(({\bf k}_1\cdot {\bf v})^2 - {\bf k}_1^2)}
\;
\frac{
\omega_2 (\omega_2 ({\bf k}_2\cdot {\bf v}) - {\bf k}_2^2)
-
{\bf k}_2^2 (\omega_2 - {\bf k}_2\cdot {\bf v})}
{{\bf k}_2^2
(\omega_2 - {\bf k}_2\cdot {\bf v})\:
(({\bf k}_2\cdot {\bf v})^2 - {\bf k}_2^2)}
\Bigg\} \;.
\ea

We now compute 
$\langle E_a^i(t_1,{\bf r}_1)E_b^j(t_2,{\bf r}_2) \rangle$ given by  \ba
\label{EiEj-x}
\langle E_a^i(t_1,{\bf r}_1) E_b^j(t_2,{\bf r}_2) \rangle
&=& \int_{-\infty +i\sigma}^{\infty +i\sigma}
{d\omega_1 \over 2\pi} \int_{-\infty +i\sigma}^{\infty +i\sigma}
{d\omega_2 \over 2\pi}
\int {d^3k_1 \over (2\pi)^3}
\int {d^3k_2 \over (2\pi)^3}
\\[2mm] \nonumber
&\times& e^{-i(\omega_1 t_1 - {\bf k}_1\cdot {\bf r}_1
+ \omega_2 t_2 - {\bf k}_2\cdot {\bf r}_2)}
\langle E_a^i(\omega_1,{\bf k}_1)
E_b^j(\omega_2,{\bf k}_2) \rangle \;.
\ea
Zeros of $(\omega_i^2\varepsilon_T(\omega_i,{\bf k}_i)- {\bf k}_i^2)$,
$(\omega_i^2\varepsilon_L(\omega_i,{\bf k}_i)$ and of 
$(\omega_i - {\bf k}_i \cdot {\bf v} +i0^+)$ with $i=1,2$ 
contribute to the integrals over $\omega_1$ and $\omega_2$.
As already mentioned, the equations 
$\omega^2\varepsilon_T(\omega,{\bf k}) - {\bf k}^2 = 0$
and $\varepsilon_L(\omega,{\bf k}) = 0$ determine, respectively, 
the transverse and longitudinal plasma modes, while
$\omega - {\bf k} \cdot {\bf v} = 0$ corresponds to the interaction 
of plasma particles of velocity ${\bf v}$ with the modes of phase
velocity ${\bf v}_\phi \equiv \omega /|{\bf k}|$. Since the system 
under consideration is stable - the collective modes are expected
to be damped, all zeros of 
$(\omega_i^2\varepsilon_T(\omega_i,{\bf k}_i)- {\bf k}_i^2)$
and $(\omega_i^2\varepsilon_L(\omega_i,{\bf k}_i)$ lie in the lower 
half-plane of complex $\omega$. Consequently, the contributions 
associated with these zeros exponentially decay in time and they 
vanish in the long-time limit of both $t_1$ and $t_2$. 

We further consider the long-time limit of 
$\langle E_a^i(t_1,{\bf r}_1) E_b^j(t_2,{\bf r}_2) \rangle$ and 
then, the only non-vanishing contribution corresponds to the 
poles at $\omega_1 = {\bf k}_1 \cdot {\bf v}$ and 
$\omega_2 = {\bf k}_2 \cdot {\bf v}$. This contribution reads
\ba
\label{EiEj-tr-1}
&&\langle E_a^i(t_1,{\bf r}_1) 
E_b^j(t_2,{\bf r}_2) \rangle_\infty
= - \frac{g^2}{2}\,\delta^{ab}
\int {d^3k_1 \over (2\pi)^3} {d^3k_2 \over (2\pi)^3} \:
(2\pi )^3\delta^{(3)}({\bf k}_1 + {\bf k}_2)
\\[2mm] \nonumber 
&\times&
\int {d^3p \over (2\pi)^3} \, f({\bf p}) \:
e^{-i(\omega_1 t_1 - {\bf k}_1\cdot {\bf r}_1
+ \omega_2 t_2 - {\bf k}_2\cdot {\bf r}_2)}
\frac{\omega_1 \omega_2}
{{{\bf k}_1^2 {\bf k}_2^2}}
\\[2mm] \nonumber
&\times \Bigg[&
\frac{\omega_1 k_1^i}{\omega_1^2 \varepsilon_L(\omega_1,{\bf k}_1)} 
+
\frac{k_1^i ({\bf k}_1\cdot {\bf v}) - v^i{\bf k}_1^2}
{\omega_1^2 \varepsilon_T(\omega_1,{\bf k}_1)-{\bf k}_1^2} 
\Bigg] 
\Bigg[
\frac{\omega_2 k_2^j}{\omega_2^2 \varepsilon_L(\omega_2,{\bf k}_2)} 
+
\frac{ v^j {\bf k}_2^2 - k_2^j ({\bf k}_2 \cdot {\bf v})}
{\omega_2^2 \varepsilon_T(\omega_2,{\bf k}_2)-{\bf k}_2^2} 
\Bigg]
\Bigg|_{\omega_1={\bf k}_1\cdot {\bf v}, \;,\;
\omega_2={\bf k}_2\cdot {\bf v}} \;.
\ea

The correlation function (\ref{EiEj-tr-1}) can be rewritten as
\ba
\langle E_a^i(t_1,{\bf r}_1) E_b^j(t_2,{\bf r}_2) \rangle_\infty 
&=& 
\int {d\omega \over 2\pi} {d^3k \over (2\pi)^3}
e^{-i \big(\omega (t_1 - t_2)
 - {\bf k}\cdot ({\bf r}_1 - {\bf r}_2)\big)} \langle E_a^i E_b^j\rangle_{\omega \, {\bf k}} \;,
\ea
where the fluctuation spectrum is
\ba
\label{EiEj-spec-1}
&&\langle E_a^i E_b^j\rangle_{\omega \, {\bf k}}
= \frac{g^2}{2}\,\delta^{ab}
\int {d^3p \over (2\pi)^3} \, f({\bf p}) \:
2\pi \delta (\omega - {\bf k} \cdot {\bf v})
\frac{\omega^2}{{\bf k}^4}
\\[2mm] \nonumber 
&\times \Bigg\{&
\frac{k^i}{\omega^2 \varepsilon_L(\omega,{\bf k})} \:
\frac{k^j}{\omega^2 \varepsilon_L(-\omega,{\bf k}^2)} \:
\omega^2
+
\frac{k^i}{\omega^2 \varepsilon_L(\omega,{\bf k})} \:
\frac{ v^j {\bf k}^2 - k^j ({\bf k} \cdot {\bf v})}
{\omega^2 \varepsilon_T(-\omega,{\bf k})-{\bf k}^2}\: \omega
\\[2mm]\nonumber
&+&
\frac{ v^i {\bf k}^2 - k^i ({\bf k} \cdot {\bf v})}
{\omega^2 \varepsilon_T(\omega,{\bf k})-{\bf k}^2}\:
\frac{k^j}{\omega^2 \varepsilon_L(-\omega,{\bf k})} \:
\omega
+
\frac{k^i ({\bf k} \cdot {\bf v}) - v^i{\bf k}^2}
{\omega^2 \varepsilon_T(\omega,{\bf k})-{\bf k}^2} \:
\frac{k^j ({\bf k}\cdot {\bf v}) - v^j{\bf k}^2} {\omega^2 \varepsilon_T(-\omega,{\bf k})-{\bf k}^2}
\Bigg\} 
\;.
\ea
One easily proves that the second and third contribution to the 
fluctuation spectrum (\ref{EiEj-spec-1}) vanish due to the plasma 
isotropy. Taking into account that for real $\omega$ and ${\bf k}$,
$\varepsilon_s(-\omega,-{\bf k})= \varepsilon_s^*(\omega,{\bf k})$
with $s=L,T$, the fluctuation spectrum (\ref{EiEj-spec-1}) can be 
written as 
\ba
\label{EiEj-spec-2}
\langle E_a^i E_b^j\rangle_{\omega \, {\bf k}}
&=& \frac{g^2}{2}\,\delta^{ab}
\int {d^3p \over (2\pi)^3} \, f ({\bf p}) \:
2\pi \delta (\omega - {\bf k} \cdot {\bf v})
\frac{\omega^2}{{\bf k}^4}
\\[2mm] \nonumber 
&\times \Bigg\{&
\frac{\omega^2 k^ik^j}{|\omega^2 \varepsilon_L(\omega,{\bf k})|^2}
+
\frac{\big( k^i ({\bf k} \cdot {\bf v}) - v^i{\bf k}^2 \big)
\big(k^j ({\bf k}\cdot {\bf v}) - v^j{\bf k}^2 \big)}
{|\omega^2 \varepsilon_T(\omega,{\bf k})-{\bf k}^2|^2}
\Bigg\} 
\;.
\ea
Due to the plasma isotropy, the expression, which enters the 
transverse contribution, can be further rewritten as 
\ba
\int {d^3p \over (2\pi)^3} \, f ({\bf p}) \:
2\pi \delta (\omega - {\bf k} \cdot {\bf v})
\big( k^i ({\bf k} \cdot {\bf v}) - v^i{\bf k}^2 \big)
\big(k^j ({\bf k}\cdot {\bf v}) - v^j{\bf k}^2 \big)
\\[2mm] \nonumber 
= \frac{1}{2}
\Big(\delta^{ij} - \frac{k^ik^j}{{\bf k}^2}\Big)
{\bf k}^2
\int {d^3p \over (2\pi)^3} \, f({\bf p}) \:
2\pi \delta (\omega - {\bf k} \cdot {\bf v}) 
\big(({\bf k}^2 {\bf v}^2 - ({\bf k}\cdot {\bf v})^2 \big) \;.
\ea

In the equilibrium plasma, the imaginary part of 
$\varepsilon_T(\omega,{\bf k})$ is given by the formula 
(\ref{Im-eT-eq}) while $\Im \varepsilon_L(\omega,{\bf k})$
found from Eq.~(\ref{eL}) by means of the identity (\ref{iden1})
equals
\be
\label{Im-eL-eq}
\Im \varepsilon_L(\omega,{\bf k}) 
= \frac{\pi g^2 \omega}{2T{\bf k}^2} 
\int {d^3p \over (2\pi)^3} \:
\delta(\omega - {\bf k}\cdot {\bf v}) \: f({\bf p})
\;.
\ee
The equilibrium fluctuation spectrum (\ref{EiEj-spec-2}) 
expressed through $\Im \varepsilon_L(\omega,{\bf k})$
and $\Im \varepsilon_T(\omega,{\bf k})$ is 
\ba
\label{EiEj-spec-FD}
\langle E_a^i E_b^j\rangle_{\omega \, {\bf k}}
= 
2 \delta^{ab} T \omega^3 \bigg[
\frac{k^ik^j}{{\bf k}^2}
\frac{\Im \varepsilon_L(\omega,{\bf k})}
{|\omega^2 \varepsilon_L(\omega,{\bf k})|^2}
+
\Big(\delta^{ij} - \frac{k^ik^j}{{\bf k}^2}\Big)
\frac{\Im \varepsilon_T(\omega,{\bf k})}
{|\omega^2 \varepsilon_T(\omega,{\bf k})-{\bf k}^2|^2}
\bigg] \;,
\ea
which for the longitudinal fields gives
\be
\label{EE-spectrum-eq}
\langle E_a^i E_b^i\rangle_{\omega {\bf k}} 
= 2 \delta^{ab} \frac{T}{\omega}
\frac{\Im \varepsilon_L(\omega,{\bf k})} 
{|\varepsilon_L(\omega,{\bf k})|^2}
\;.
\ee
The electromagnetic counterpart of Eq.~(\ref{EiEj-spec-FD}),
which is derived in \cite{Mrow2007}, agrees with Eq.~(11.2.2.6) 
from \cite{Akh75} provided by the fluctuation-dissipation 
relation. In Sec.~\ref{subsec-statistics} we show that 
Eqs.~(\ref{EiEj-spec-FD},\ref{EE-spectrum-eq}) are still valid
when quarks and gluons obey quantum statistics but the equilibrium
formulas of $\Im \varepsilon_L(\omega,{\bf k})$ and 
$\Im \varepsilon_T(\omega,{\bf k})$ require a modification.


\section{Longitudinal chromoelectric field in 
the two-stream system}
\label{sec-2-streams}


Nonequlibrium calculations are much more difficult than the 
equilibrium ones. The first problem is to invert the matrix
$\Sigma^{ij}(\omega,{\bf k})$ defined by Eq.~(\ref{matrix-sigma}).
In the case of longitudinal electric field, which is discussed
here, it is solved trivially. We start with Eq.~(\ref{E-field2})
projecting it on ${\bf k}$ and assuming that ${\bf E}$ and ${\bf E}_0$ 
are purely longitudinal. Then, the matrix (\ref{matrix-sigma}) is
replaced by the scalar function. 

Further, we neglect the first term in the r.h.s. of Eq.~(\ref{E-field2}). 
This term vanishes in isotropic systems; it is of order $g^2$ higher  than the second term; it is also expected to be small in nonrelativistic 
regime due to the smallness of particle velocity. So, there are good 
reasons to neglect it. Eliminating ${\bf E}_0$ by means of the first 
Maxwell equation we obtain Eq.~(\ref{E-field-final-iso-long2}) which 
was previously derived for the case of isotropic plasma. In the 
following we consider fluctuations of longitudinal chromoelectric 
fields in the two-stream system. A nonrelativistic approximation 
was adopted to discuss this unstable system in the paper \cite{Mrow2007} 
where electromagnetic plasmas were studied. Our considerations here
are fully relativistic as the nonrelativistic approximation is
usually irrelevant for the quark-gluon plasma.

The distribution function of the two-stream system is chosen to be 
\be
\label{f-2-streams}
f({\bf p}) = (2\pi )^3 n 
\Big[\delta^{(3)}({\bf p} - {\bf q}) + \delta^{(3)}({\bf p} + {\bf q}) \Big] \;,
\ee
where $n$ is the effective parton density in a single stream. The 
distribution function (\ref{f-2-streams}) should be treated as an 
idealization of the two-peak distribution where the particles have 
momenta close to ${\bf q}$ or $-{\bf q}$ but it is not required that 
the momenta are exactly ${\bf q}$ or $-{\bf q}$. There is no Bose 
condensation of gluons which would invalidate our derivation of the 
correlation function of distribution functions (\ref{corr-final-quant})
due to non-trivial bosonic correlations.

To compute $\varepsilon_L(\omega,{\bf k})$ we first perform integration 
by parts in Eq.~(\ref{eL}) and then, substituting the distribution 
function (\ref{f-2-streams}) into the resulting formula, we obtain 
\ba
\label{eL-3}
\varepsilon_L(\omega,{\bf k}) 
= 1 - \mu^2 
\frac{{\bf k}^2 -({\bf k} \cdot {\bf u})^2}{{\bf k}^2}
\bigg[
  \frac{1}{(\omega - {\bf k} \cdot {\bf u})^2}
+ \frac{1}{(\omega + {\bf k} \cdot {\bf u})^2}
\bigg]
\\[2mm] \nonumber
= \frac{\big(\omega - \omega_+({\bf k})\big)
\big(\omega + \omega_+({\bf k})\big)
\big(\omega - \omega_-({\bf k})\big)
\big(\omega + \omega_-({\bf k})\big)}
{\big(\omega^2 - ({\bf k} \cdot {\bf u})^2\big)^2} 
\;,
\ea
where ${\bf u} \equiv {\bf q}/E_{\bf q}$ is the stream velocity, 
$\mu^2 \equiv g^2n/2 E_{\bf q}$ and $\pm \omega_{\pm}({\bf k})$
are the four roots of the dispersion equation 
$\varepsilon_L(\omega,{\bf k}) = 0$ which read 
\be
\label{roots}
\omega_{\pm}^2({\bf k}) = \frac{1}{{\bf k}^2}
\bigg[{\bf k}^2 ({\bf k} \cdot {\bf u})^2
+ \mu^2 \big({\bf k}^2 - ({\bf k} \cdot {\bf u})^2\big)
\pm \mu \sqrt{\big({\bf k}^2 - ({\bf k} \cdot {\bf u})^2\big)
\Big(4{\bf k}^2 ({\bf k} \cdot {\bf u})^2 +
\mu^2 \big({\bf k}^2 - ({\bf k} \cdot {\bf u})^2\big)\Big)} 
\; \bigg] \;.
\ee
One shows that $0 < \omega_+({\bf k}) \in R$ for any ${\bf k}$ 
while $\omega_-({\bf k})$ is imaginary for 
${\bf k}^2 ({\bf k} \cdot {\bf u})^2 
< 2 \mu^2 \big({\bf k}^2 - ({\bf k} \cdot {\bf u})^2\big)$
when it represents the well-known two-stream electrostatic 
instability generated due to the mechanism analogous to
the Landau damping. For ${\bf k}^2 ({\bf k} \cdot {\bf u})^2 
\ge 2 \mu^2 \big({\bf k}^2 - ({\bf k} \cdot {\bf u})^2\big)$, 
the mode is stable, $0 < \omega_-({\bf k}) \in R$. 

When the chromoelectric fields are purely longitudinal and 
the first term in the r.h.s. of Eq.~(\ref{E-field2}) is neglected, the 
correlation function 
$\langle E_a^i(\omega_1,{\bf k}_1) E_b^i(\omega_2,{\bf k}_2) \rangle$
is given by the first term of Eq.~(\ref{EiEj-1}) as
\ba
\label{EiEi-L}
\langle E_a^i(\omega_1,{\bf k}_1)  E_b^i(\omega_2,{\bf k}_2) \rangle
= \frac{g^2}{2}\,\delta^{ab} 
(2\pi )^3\delta^{(3)}({\bf k}_1 + {\bf k}_2)
\frac{{\bf k}_1 \cdot {\bf k}_2}{{\bf k}_1^2 {\bf k}_2^2}
\frac{1}{\varepsilon_L(\omega_1,{\bf k}_1)} \:
\frac{1}{\varepsilon_L(\omega_2,{\bf k}_2)} \:
\\[2mm]\nonumber
\times
\int {d^3p \over (2\pi)^3} \, 
\frac{f({\bf p})}
{(\omega_1 - {\bf k}_1\cdot {\bf v}) \:
(\omega_2 - {\bf k}_2\cdot {\bf v})} \;.
\ea
Substituting the distribution function (\ref{f-2-streams})
and the dielectric function (\ref{eL-3}) into Eq.~(\ref{EiEi-L}),
one finds
\ba
\label{E-fluc-2-stream-2} 
\langle E_a^i(\omega_1,{\bf k}_1) 
E_b^i(\omega_2,{\bf k}_2) \rangle 
&=& - g^2 \delta^{ab} n \: 
\frac{(2\pi)^3\delta^{(3)}({\bf k}_1 + {\bf k}_2)}
{{\bf k}_1^2} \Big[\omega_1 \omega_2 + 
({\bf k}_1 \cdot {\bf u})({\bf k}_2 \cdot {\bf u}) \Big]
\\ [2mm] \nonumber 
&\times&
\frac{\omega_1^2 - ({\bf k}_1 \cdot {\bf u})^2} 
{\big(\omega_1 - \omega_-({\bf k}_1)\big)
\big(\omega_1 + \omega_-({\bf k}_1)\big)
\big(\omega_1 - \omega_+({\bf k}_1)\big)
\big(\omega_1 + \omega_+({\bf k}_1)\big) }
\\ [2mm] \nonumber 
&\times&
\frac{\omega_2^2 - ({\bf k}_2 \cdot {\bf u})^2} 
{\big(\omega_2 - \omega_-({\bf k}_2)\big)
\big(\omega_2 + \omega_-({\bf k}_2)\big)
\big(\omega_2 - \omega_+({\bf k}_2)\big)
\big(\omega_2 + \omega_+({\bf k}_2)\big) }
\;.
\ea
One observes that the poles of the correlation function
$\langle E_a^i(\omega_1,{\bf k}_1) 
E_b^i(\omega_2,{\bf k}_2) \rangle$ 
at $\omega_1 = {\bf k}_1 {\bf v}$ and 
$\omega_2 = {\bf k}_2 {\bf v}$, which give the stationary 
contribution to the equilibrium fluctuation spectrum, have 
disappeared in Eq.~(\ref{E-fluc-2-stream-2}) as the
inverse dielectric functions vanish at these points.

The correlation function 
$\langle E_a^i(t_1,{\bf r}_1)E_b^i(t_2,{\bf r}_2) \rangle$ 
is given by Eq.~(\ref{EiEj-x}) with  $\langle E_a^i(\omega_1,{\bf k}_1) 
E_b^i(\omega_2,{\bf k}_2) \rangle$ defined by 
Eq.~(\ref{E-fluc-2-stream-2}). Performing the trivial 
integration over ${\bf k}_2$ and taking into account
that $\omega_{\pm}(-{\bf k}) = \omega_{\pm}({\bf k})$,
one finds
\ba
\nonumber
\langle E_a^i(t_1,{\bf r}_1) E_b^i(t_2,{\bf r}_2) \rangle
&=& g^2 \delta^{ab} n 
\int_{-\infty +i\sigma}^{\infty +i\sigma}
{d\omega_1 \over 2\pi i}
\int_{-\infty +i\sigma}^{\infty +i\sigma}
{d\omega_2 \over 2\pi i}
\int {d^3k \over (2\pi)^3}
\frac{e^{-i \big(\omega_1 t_1 + \omega_2 t_2
- {\bf k}({\bf r}_1 - {\bf r}_2)\big)}}
{{\bf k}^2}
\big[\omega_1 \omega_2 - ({\bf k} \cdot {\bf u})^2 \big]
\\ [2mm] \nonumber 
&\times&
\frac{\omega_1^2 - ({\bf k} \cdot {\bf u})^2} 
{\big(\omega_1 - \omega_-({\bf k})\big)
\big(\omega_1 + \omega_-({\bf k})\big)
\big(\omega_1 - \omega_+({\bf k})\big)
\big(\omega_1 + \omega_+({\bf k})\big) }
\\ [2mm] 
\label{EL-fluc-x-stream1}
&\times&
\frac{\omega_2^2 - ({\bf k} \cdot {\bf u})^2} 
{\big(\omega_2 - \omega_-({\bf k})\big)
\big(\omega_2 + \omega_-({\bf k})\big)
\big(\omega_2 - \omega_+({\bf k})\big)
\big(\omega_2 + \omega_+({\bf k})\big)}
\;.
\ea
There are 16 contributions to the integrals over $\omega_1$ 
and $\omega_2$ in Eq.~(\ref{EL-fluc-x-stream1}) 
related to the poles at $\pm \omega_{\pm}$. Summing up the 
contributions, we get after lengthy calculation
\ba
\label{EL-fluc-x-stream6}
\langle E_a^i(t_1,{\bf r}_1) E_b^i(t_2,{\bf r}_2) \rangle
&=& \frac{g^2}{2}\,\delta^{ab} n  \int {d^3k \over (2\pi)^3} 
\frac{e^{i {\bf k}({\bf r}_1 - {\bf r}_2)}}{{\bf k}^2}
\frac{1}{(\omega_+^2 - \omega_-^2)^2}
\\[2mm] \nonumber
\times 
\bigg\{
\frac{\big(\omega_+^2 - ({\bf k} \cdot {\bf u})^2\big)^2}
{\omega_+^2}
&\Big[&
\big(\omega_+^2 - ({\bf k} \cdot {\bf u})^2\big)
\cos \big( \omega_+ (t_1 + t_2)\big)
+ 
\big(\omega_+^2 + ({\bf k} \cdot {\bf u})^2\big)
\cos \big(\omega_+ (t_1 - t_2)\big)
\Big]
\\[2mm] \nonumber
- 
\frac{
\big(\omega_+^2 - ({\bf k} \cdot {\bf u})^2\big)
\big(\omega_-^2 - ({\bf k} \cdot {\bf u})^2\big)} 
{\omega_+ \omega_- }
&\Big[&
\big(\omega_+ \omega_- - ({\bf k} \cdot {\bf u})^2\big)
\cos (\omega_+ t_1 + \omega_- t_2)
+
\big(\omega_+ \omega_- + ({\bf k} \cdot {\bf u})^2\big)
\cos (\omega_+ t_1 - \omega_- t_2)
\\[2mm] \nonumber
&+& 
\big(\omega_+ \omega_- - ({\bf k} \cdot {\bf u})^2\big)
\cos (\omega_- t_1 + \omega_+ t_2)
+
\big(\omega_+ \omega_- + ({\bf k} \cdot {\bf u})^2\big)
\cos (\omega_- t_1 - \omega_+ t_2)
\Big]
\\[2mm] \nonumber
+
\frac{
\big(\omega_-^2 - ({\bf k} \cdot {\bf u})^2\big)^2} 
{\omega_-^2}
&\Big[&
\big(\omega_-^2 - ({\bf k} \cdot {\bf u})^2\big)
\cos \big(\omega_- (t_1 + t_2)\big)
+
\big(\omega_-^2 + ({\bf k} \cdot {\bf u})^2\big)
\cos \big(\omega_- (t_1 - t_2)\big)
\Big]
\bigg\}\;.
\ea

Let us now consider the domain of wave vectors
obeying ${\bf k}^2 ({\bf k} \cdot {\bf u})^2 
< 2 \mu^2 \big({\bf k}^2 - ({\bf k} \cdot {\bf u})^2\big)$
when $\omega_-({\bf k})$ is imaginary and it represents 
the unstable electrostatic mode. We write down 
$\omega_-({\bf k})$ given by Eq.~(\ref{roots}) as $i \gamma_{\bf k}$ 
with $0 < \gamma_{\bf k} \in R$. We are interested in the 
contributions to the correlation function coming from the 
unstable modes. The contributions, which are the fastest growing 
functions of $(t_1+t_2)$ and $(t_1-t_2)$, correspond to the last 
term in Eq.~(\ref{EL-fluc-x-stream6}). 
The contributions provide
\ba
\label{EL-fluc-x-stream7}
\langle E_a^i(t_1,{\bf r}_1) E_b^i(t_2,{\bf r}_2) 
\rangle_{\rm unstable}
&=& \frac{g^2}{2}\,\delta^{ab} n 
\int {d^3k \over (2\pi)^3} 
\frac{e^{i {\bf k}({\bf r}_1 - {\bf r}_2)}}{{\bf k}^2}
\frac{1}{(\omega_+^2 - \omega_-^2)^2}
\frac{
\big(\gamma_{\bf k}^2 + ({\bf k} \cdot {\bf u})^2\big)^2} 
{\gamma_{\bf k}^2}
\\[2mm] \nonumber
&\times& 
\Big[
\big(\gamma_{\bf k}^2 + ({\bf k} \cdot {\bf u})^2\big)
\cosh \big(\gamma_{\bf k} (t_1 + t_2)\big)
+
\big(\gamma_{\bf k}^2 - ({\bf k} \cdot {\bf u})^2\big)
\cosh \big(\gamma_{\bf k} (t_1 - t_2)\big) \Big] \;.
\ea
As seen, the correlation function (\ref{EL-fluc-x-stream7}) is
invariant with respect to space translations -- it depends on the 
difference $({\bf r}_1 - {\bf r}_2)$ only. The initial plasma state
is on average homogeneous and it remains like this in course of
the system's temporal evolution. The time dependence of the correlation 
function (\ref{EL-fluc-x-stream7}) is very different from the space
dependence. The electric fields exponentially grow and so does 
the correlation function both in $(t_1 + t_2)$ and $(t_1 - t_2)$.
The fluctuation spectrum also evolves in time as the growth rate
of unstable modes is wave-vector dependent. After a sufficiently
long time the fluctuation spectrum is dominated by the fastest 
growing modes.


\section{Discussion and Outlook}
\label{sec-discussion}


We discuss here validity of the results obtained in the previous 
sections and their possible applications. We also briefly consider
fluctuations of color charges and currents, and finally we summarize 
our study. We start with the important problem of gauge dependence 
of the correlation functions.

\subsection{Gauge dependence of the correlation functions}
\label{subsec-gauge}

The linearized transport and Yang-Mills equations, which are solved
in Sec.~\ref{sec-initial-value}, are not gauge covariant, and thus
the question arises how the correlation functions derived in
Secs.~\ref{sec-fluc-B} and \ref{sec-fluc-E} depend on a gauge. 
We consider the functions like $\langle H_a (x_1) K_b(x_2) \rangle$
where $x_1 \equiv (t_1,{\bf r}_1)$ and $x_2 \equiv (t_2,{\bf r}_2)$ 
are four-positions and $H_a (x)$ and $K_b(x)$ are the fields belonging 
to the adjoint representation of ${\rm SU}(N_c)$ group which transform 
under infinitesimal gauge transformations as
\be
\label{gauge-trans}
H_a (x) \rightarrow H_a (x) + f_{abc} \lambda_b(x) H_c (x) \;,
\ee
where $f_{abc}$ is the structure constant of ${\rm SU}(N_c)$ and
$\lambda_a(x)$ is the infinitesimal gauge parameter. 
The correlation function $\langle H_a (x_1) K_b(x_2) \rangle$
transforms under the gauge transformation (\ref{gauge-trans}) as
\be
\label{corr-fun-trans}
\langle H_a (x_1) K_b(x_2) \rangle 
\rightarrow
\langle H_a (x_1) K_b(x_2) \rangle  + f_{acd} \lambda_c(x_1) \langle H_d (x_1) K_b(x_2) \rangle
+ f_{bef} \lambda_e(x_2) \langle H_a (x_1) K_f(x_2) \rangle \;.
\ee

We consider in this paper the fluctuations around a colorless 
sate, and consequently the correlation functions derived in
Secs.~\ref{sec-fluc-B} and \ref{sec-fluc-E} have a very simple
color structure. Namely,
\be
\label{corr-fun-struc}
\langle H_a (x_1) K_b(x_2) \rangle
= \delta^{ab} L (x_1,x_2) \;.
\ee
Then, the transformation law (\ref{corr-fun-trans}) gives
\be
\label{corr-fun-trans2}
\delta^{ab} L (x_1,x_2)
\rightarrow
\big( \delta^{ab} 
+ f_{acb} \lambda_c(x_1) 
+ f_{bea} \lambda_e(x_2) \big) L (x_1,x_2)
\;.
\ee
One observes that with the transformation (\ref{corr-fun-trans2}),
the correlation function $\langle H_a (x_1) K_a(x_2) \rangle
= (N_c^2-1) L(x_1,x_2)$ is gauge invariant (due to the antisymmetry 
of the structure constants), even so the function is not local
in coordinate space. 

We conclude this section by saying that the correlation functions, which 
are discussed in this paper, are gauge invariant after the trivial sum 
over colors is taken. This happens because only small fluctuations
around colorless state are considered. 

\subsection{Effect of quantum statistics}
\label{subsec-statistics}

Deriving the correlation functions, we have assumed that quarks
and gluons obey Boltzmann statistics but the effect of quantum 
statistics can be easily taken into account. Instead of 
Eqs.~(\ref{dQ-dQ}, \ref{dbarQ-dbarQ}, \ref{dG-dG}),
the free correlation function (\ref{corr-final-quant}), which is 
obtained in the Appendix, suggests
\ba
\label{dQ-dQ-qs}
\langle \delta Q^{mn}(t_1,{\bf r}_1,{\bf p}_1) 
\delta Q^{pr}(t_2,{\bf r}_2,{\bf p}_2)\rangle_{\rm free}
= \delta^{mr} \delta^{np} 
(2\pi )^3 \delta^{(3)}({\bf p}_1 - {\bf p}_2) \,
\delta^{(3)}\big({\bf r}_2 - {\bf r}_1 
- {\bf v}_1(t_2 - t_1)\big) \: n({\bf p}_1) 
\big(1 - n({\bf p}_1) \big)\;,
\\[2mm] 
\label{dbarQ-dbarQ-qs}
\langle \delta \bar Q^{mn}(t_1,{\bf r}_1,{\bf p}_1) 
\delta \bar Q^{pr}(t_2,{\bf r}_2,{\bf p}_2)\rangle_{\rm free}
= \delta^{mr} \delta^{np} 
(2\pi )^3 \delta^{(3)}({\bf p}_1 - {\bf p}_2) \,
\delta^{(3)}\big({\bf r}_2 - {\bf r}_1 
- {\bf v}_1(t_2 - t_1)\big) \: \bar n({\bf p}_1) 
\big(1 - \bar n({\bf p}_1) \big)  \;,
\\[2mm] 
\label{dG-dG-qs}
\langle \delta G^{ab}(t_1,{\bf r}_1,{\bf p}_1) 
\delta G^{cd}(t_2,{\bf r}_2,{\bf p}_2)\rangle_{\rm free}
= \delta^{ad} \delta^{bc} 
(2\pi )^3 \delta^{(3)}({\bf p}_1 - {\bf p}_2) \,
\delta^{(3)}\big({\bf r}_2 - {\bf r}_1 
- {\bf v}_1(t_2 - t_1)\big) \: n_g({\bf p}_1) 
\big(1 + n_g({\bf p}_1) \big)\;.
\ea

With the initial correlations given by 
Eq.~(\ref{dQ-dQ-qs}, \ref{dbarQ-dbarQ-qs}, \ref{dG-dG-qs}), 
the correlation functions derived 
Secs.~\ref{sec-fluc-B}, \ref{sec-fluc-E}, \ref{sec-2-streams} 
are somewhat modified. Instead of the effective distribution 
function $f({\bf p})$, there are two different effective 
distribution functions $f({\bf p})$ and $\tilde f({\bf p})$. 
The function, which enters the dielectric tensor
(except Eqs.~(\ref{Im-eT-eq}, \ref{Im-eL-eq})), is, as previously,  
$f({\bf p}) \equiv n({\bf p}) + \bar n ({\bf p}) + 2N_c n_g({\bf p})$ 
but the function originating from the initial correlation functions 
equals  
$\tilde f({\bf p}) \equiv n({\bf p}) \big(1 - n({\bf p}) \big) 
+ \bar n ({\bf p}) \big(1 - \bar n({\bf p}) \big) 
+ 2N_c n_g({\bf p})\big(1 + n_g({\bf p}) \big)$. 
The equilibrium formulas of $\Im \varepsilon_T$ (\ref{Im-eT-eq})
and $\Im \varepsilon_L$ (\ref{Im-eL-eq}) are expressed through
$\tilde f({\bf p})$ not $f({\bf p})$, and consequently the final 
fluctuation-dissipation relations 
(\ref{BiBj-spec-eq}, \ref{EiEj-spec-FD}, \ref{EE-spectrum-eq})
remain unchanged.

\subsection{Validity of linear collisionless approach}
\label{subsec-linear}

We first note that the approach adopted here is dynamically equivalent 
to the Hard Loop approximation which is commonly applied to 
equilibrium quark-gluon plasma (for a review see \cite{Thoma:1995ju}) 
but has been extended to nonequilibrium systems as well 
\cite{Mrowczynski:2000ed,Mrowczynski:2004kv}. The approximation,
which can be formulated in terms of resumed diagrams or transport
theory, allows one to study soft Abelian or nonAbelian fields of small 
amplitude in the background of hard particles. Below we discuss in 
more detail specific steps of our derivation of the fluctuation spectra.

We have started with the Yang-Mills and collisionless transport equations. 
The collisionless approximation is justified for the time scales which 
are much shorter than those of collisional processes. As discussed in 
\cite{Arnold:1998cy}, the characteristic inverse time of system's 
evolution due to inter-parton collisions is 
$t_{\rm hard}^{-1} \sim g^4 {\ln}(1/g)T$ or 
$t_{\rm soft}^{-1} \sim g^2 {\ln}(1/g)T$, depending whether the 
momentum transfer in a collision is of order $T$ or $gT$ with $T$ being 
a typical parton momentum ($T$ is the temperature in the equilibrium 
plasma). Since an evolution of color degrees of freedom is due to the 
soft collisions \cite{Arnold:1998cy}, the correlation functions derived 
in this paper are valid for time intervals shorter that $t_{\rm soft}$.

Another time scale limitation comes from the fact that performing
the linearization of equations of motion, the state, that small
fluctuations around it are considered, is assumed to be stationary.
Except the equilibrium state or a state kept stationary by external
conditions, nonequilibrium states evolve in time. Therefore our 
approach is valid for the time scales which are much shorter than 
a characteristic time of evolution of the whole system. In equilibrium,
the latter time is infinite and there is no limitation. 
 Performing the linearization procedure, we have assumed that 
$|Q^0| \gg |\delta Q|$ and the quadratic terms in $\delta Q$ or $A^\mu$ 
have been neglected. Estimating $\delta Q$, which is given by 
Eq.~(\ref{trans-eq-lin}), in the following way
\be
\delta Q(t,{\bf r},{\bf p}) \sim g \int_0^t dt' \big({\bf E}(t',{\bf r}) 
+ {\bf v} \times {\bf B}(t',{\bf r})\big) \nabla_p n({\bf p})
\sim g \, t \, H \, \frac{n}{T} \;, 
\ee
where $H$ is the magnitude of ${\bf E}$ or ${\bf B}$, the 
condition $n \gg |\delta Q|$ provides $T \gg g \, t \, H$.

The assumptions discussed above can be quantitatively checked only 
for a well defined plasma state under consideration. Qualitatively,
the method presented here is limited to small amplitude fluctuations 
which are observed for a sufficiently short interval of time. 

\subsection{Fluctuations of other chromodynamic quantities}
\label{subsec-other}

We have studied in the previous sections fluctuations of 
chromoelectric and chromomagnetic fields but fluctuations 
of other quantities can be inferred from the presented 
formulas. For example, let us consider fluctuations of color 
charges as given by the correlation function 
$\langle \rho_a(\omega_1,{\bf k}_1) \rho_b(\omega_2,{\bf k}_2) \rangle$
\footnote{Fluctuations of color charges and currents in the system 
of free quarks and gluons have been earlier discussed in
\cite{Mrowczynski:1996vh}. Unfortunately, some numerical coefficients
are incorrect in \cite{Mrowczynski:1996vh}. Specifically, the coefficient
should 1/2 instead of 1/8 in Eqs.~(3, 4), while in the unnumbered equation
following Eq.~(4) there should be 2 instead of $1/2\pi$.}.
Using the first Maxwell equations (\ref{Maxwell-eqs-k}), one 
immediately finds
\be
\langle \rho_a(\omega_1,{\bf k}_1) \rho_b(\omega_2,{\bf k}_2) \rangle
= - k_1^i k_2^j \langle E_a^i(\omega_1,{\bf k}_1) E_b^j(\omega_2,{\bf k}_2) 
\rangle \;.
\ee 
Then, Eq.~(\ref{EiEj-spec-2}) provides the spectrum of color charge
fluctuations in the isotropic plasma
\ba
\label{rho-rho-spec}
\langle \rho_a \rho_b \rangle_{\omega \, {\bf k}}
= \frac{g^2}{2}\,\delta^{ab} 
\int {d^3p \over (2\pi)^3} \, f ({\bf p}) \:
\frac{2\pi \delta (\omega - {\bf k} \cdot {\bf v})}
{|\varepsilon_L(\omega,{\bf k})|^2} 
= 2 \delta^{ab} \frac{{\bf k}^2 T}{\omega}
\frac{\Im \varepsilon_L(\omega,{\bf k})} 
{|\varepsilon_L(\omega,{\bf k})|^2}
\;. \ea The last equality holds for the equilibrium plasma. 

Fluctuations of color currents  $\langle j_a^i(\omega_1,{\bf k}_1) j_b^j(\omega_2,{\bf k}_2) \rangle$
can be obtained in a way similar to that $\langle E_a^i(\omega_1,{\bf k}_1) 
E_b^j(\omega_2,{\bf k}_2) \rangle$ has been obtained. In the case of 
the stable system, when the initial fluctuations are forgotten, the 
spectrum of color current fluctuations can be found as
\be
\label{jj-1}
\langle j_a^i j_b^j \rangle_{\omega,{\bf k}} =
\frac{1}{\omega^2} 
\big[(\omega^2 - {\bf k}^2) \, \delta^{ik} + k^ik^k \big] 
\big[(\omega^2 - {\bf k}^2) \, \delta^{jl} + k^jk^l \big] 
\langle E_a^k E_b^l \rangle_{\omega,{\bf k}} \;.
\ee
Substituting Eq.~(\ref{EiEj-spec-FD}) into the formula (\ref{jj-1}),
one obtains the equilibrium spectrum of color current fluctuations
\ba
\label{jj-spec-eq}
\langle j_a^i j_b^j \rangle_{\omega,{\bf k}} 
= 
2 \delta^{ab} T \omega 
\bigg[
\frac{k^ik^j}{{\bf k}^2}
\frac{\Im \varepsilon_L(\omega,{\bf k})}
{|\varepsilon_L(\omega,{\bf k})|^2}
+ \frac{(\omega^2 - {\bf k}^2)^2}{\omega^4}
\Big(\delta^{ij} - \frac{k^ik^j}{{\bf k}^2}\Big)
\frac{\Im \varepsilon_T(\omega,{\bf k})}
{|\varepsilon_T(\omega,{\bf k})-{\bf k}^2/\omega^2|^2}
\bigg] 
\;.
\ea
As seen, the equilibrium spectra (\ref{rho-rho-spec}, \ref{jj-spec-eq}) 
obey the relation $\omega^2 \langle \rho_a \rho_b \rangle_{\omega \, {\bf k}} 
= k^ik^j\langle j_a^i j_b^j \rangle_{\omega,{\bf k}}$ which follows from
the (linearized) color charge conservation.

\subsection{Summary and Outlook}
\label{subsec-summary}

The calculations presented here show how to obtain spectra of 
chromodynamic fluctuations in equilibrium or nonequilibrium QGP 
as a solution of initial value problem. We first linearize the 
transport equations around the state which is on average coloreless, 
stationary and homogenous. The linearized transport equations are 
solved together with the Maxwell equations by means of the one-sided 
Fourier transformation. The time dependent fluctuation spectrum is 
expressed through the fluctuations in the initial state. The 
chromodynamic initial fluctuations are determined by the initial 
fluctuations of the distribution function. The later are identified 
with the fluctuations in a classical system of noninteracting partons. 
We compute fluctuation spectra of chromomagnetic and chromoelectric
fields in isotropic plasma. Our equilibrium results can be interpreted
as the fluctuation-dissipation relations. However, the method adopted 
here clearly shows how the system looses its memory and how the stationary 
equilibrium spectrum of fluctuations emerges. As an example of unstable 
systems, the fluctuations of longitudinal electric field in the two-stream
system are considered. The fluctuation spectrum appears to be qualitatively 
different than that of the equilibrium plasma - the collective unstable
mode does not exponentially decays but it grows and dominates the spectrum.   

The scheme of calculation, which is worked our here in detail, can 
be applied to a variety of plasma nonequilibrium states. We plan 
to compute a spectrum of chromomaganetic fluctuations in QGP 
produced at the early stage of relativistic heavy-ion collisions.
The spectrum is of particular interests as it controls transport 
properties of QGP \cite{Asakawa:2006jn}. It should be remembered,
however, that our approach, which is based on the linearized
equations of motion, deals with the quasicolorless plasma 
- the color perturbations are assumed to be small. The fluctuation 
spectrum of chromomagnetic fields in the plasma at later stages of
instability development, when the chromodynamic fields are sizeable, 
needs another treatment. At present such a spectrum is accessible 
only through numerical simulations  
\cite{Arnold:2005vb,Arnold:2005ef,Rebhan:2005re,Dumitru:2006pz,Romatschke:2006nk,Bodeker:2007fw,Berges:2007re}. 

\begin{acknowledgments}

I am grateful to Cristina Manuel and Peter Arnold for helpful 
correspondence. This work was partially supported by the Virtual 
Institute VH-VI-146 of Helmholtz Gemeinschaft.

\end{acknowledgments}

\appendix

\section{}


We compute here correlations of the distribution functions 
of free quarks or gluons using an apparatus of Quantum Field Theory
in the Keldysh-Schwinger framework which is applicable to equilibrium
and nonequilibrium systems. Actually, we do not need a whole machinery 
of the formalism but we mostly refer to it to carefully perform the Wick 
decomposition of an expectation value of product of field operators. 
For simplicity, we consider not the quark and gluon fields of QCD but 
the scalar complex field $\phi_i(x)$ with an internal degree of freedom 
labeled by the index $i$ which is further identified with color.  

As discussed in detail in {\it e.g.} \cite{Mrowczynski:1992hq}, 
the average distribution functions of particles and antiparticles
described by the field $\phi_i(x)$ are obtained from the Green's
functions
\ba
i\Delta^>_{ij}(x_1,x_2) \equiv 
\langle \phi_i(x_1) \phi^*_j(x_2) \rangle \;,
\\[2mm]
i\Delta^<_{ij}(x_1,x_2) \equiv
\langle \phi^*_j(x_2) \phi_i(x_1) \rangle \;.
\ea 
After performing the Wigner transformation
\be
\Delta (X,p) = \int d^4u \: e^{ipu} 
\Delta \Big(X+\frac{u}{2},X-\frac{u}{2}\Big) \;,
\ee 
one defines the average distribution function of particles
$f_{ij}(X,{\bf p})$ and of antiparticles $\bar f_{ij}(X,{\bf p})$,
which are on mass-shell, as
\ba
i\Delta^<_{ij}(X,p) \equiv \frac{\pi}{E_{\bf p}} \delta (E_{\bf p}-p^0)
f_{ij}(X,{\bf p}) \;,
\\[2mm]
i\Delta^>_{ij}(X,p) \equiv \frac{\pi}{E_{\bf p}} \delta (E_{\bf p}+p^0)
\bar f_{ij}(X,-{\bf p}) \;.
\ea
Taking into account the commutation relations obeyed by the
field operators, one finds that
\ba
\label{<-f}
i\Delta^<_{ij}(X,p) = \frac{\pi}{E_{\bf p}} \delta (E_{\bf p}-p^0)
f_{ij}(X,{\bf p}) 
+ \frac{\pi}{E_{\bf p}} \delta (E_{\bf p}+p^0)
\big[ \bar f_{ij}(X,-{\bf p}) + \delta^{ij} \big] \;,
\\[2mm]
\label{>-f}
i\Delta^>_{ij}(X,p) = 
 \frac{\pi}{E_{\bf p}} \delta (E_{\bf p}-p^0)
\big[ f_{ij}(X,{\bf p}) + \delta^{ij} \big]
+ \frac{\pi}{E_{\bf p}} \delta (E_{\bf p}+p^0)
\bar f_{ij}(X,-{\bf p}) \;.
\ea

\begin{figure}[t]
\includegraphics*[width=100mm]{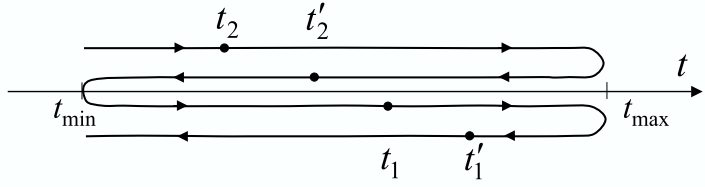}
\caption{The contour in the complex time which is used to calculate
correlations of distribution functions.
\label{fig-contour}} 
\end{figure}

We define 
\be 
\delta {\cal F}_{ij}(X,{\bf p})
\equiv 
{\cal F}_{ij}(X,{\bf p}) 
- \langle {\cal F}_{ij}(X,{\bf p}) \rangle \;,
\ee
where ${\cal F}_{ij}(X,{\bf p})$ is a microscopic (nonaveraged)
distribution function and $\langle {\cal F}_{ij}(X,{\bf p}) \rangle
= f_{ij}(X,{\bf p})$. We are interested in the correlation function 
$\langle \delta {\cal F}_{ij}(X_1,{\bf p}_1)\:
\delta {\cal F}_{kl}(X_2,{\bf p}_2) \rangle $ which is expressed 
through the fields operators as 
\ba
\label{corr-fields}
\langle \delta {\cal F}_{ij}(X_1,{\bf p}_1)\:
\delta {\cal F}_{kl}(X_2,{\bf p}_2) \rangle 
&=&
4E_{{\bf p}_1} E_{{\bf p}_2} \int \frac{dp^0_1}{2\pi}\, \Theta(p_1^0)
\int \frac{dp^0_1}{2\pi} \, \Theta(p_2^0)
\int d^4u_1 \int d^4u_1 \: e^{i(p_1 u_1 + p_2 u_2)} 
\\[2mm] \nonumber
&\times& W_{ijkl}\Big(X_1+\frac{u_1}{2},X_1-\frac{u_1}{2},
X_2+\frac{u_2}{2},X_2-\frac{u_2}{2}\Big) \;,
\ea
where
\be
W_{ijkl}(x_1,x_1',x_2,x_2')\equiv
\langle \phi^*_j(x_1') \phi_i(x_1) \phi^*_l(x_2') \phi_k(x_2) \rangle
- \langle \phi^*_j(x_1') \phi_i(x_1)\rangle
\langle \phi^*_l(x_2') \phi_k(x_2)\rangle \;. 
\ee 
The Wick theorem allows one to express an expectation value of product 
of field operators as a sum of products of expectation values of products 
of two operators. However, the theorem deals with chronologically 
ordered products of field operators. To compute the expectation value 
of any order of operators in the product irrespective of the values 
of times as in  $\langle \phi^*_j(x_1') \phi_i(x_1) \phi^*_l(x_2') \phi_k(x_2) \rangle$,
one may use contours (in the space of complex time) which run many times 
forward and backward in time as discussed in \cite{Danielewicz90}.    
We compute the expectation value
$\langle \phi^*_j(x_1') \phi_i(x_1) \phi^*_l(x_2') \phi_k(x_2) \rangle$, 
using the contour shown in Fig.~\ref{fig-contour} where the four 
branches of the contour are infinitely close to the axis of real
time and $t_{\rm min} \rightarrow -\infty$ and  $t_{\rm max} \rightarrow \infty$. Locating the time arguments on 
the contour as shown in Fig.~\ref{fig-contour}, we can formally replace 
$\langle \phi^*_j(x_1') \phi_i(x_1) \phi^*_l(x_2') \phi_k(x_2) \rangle$
by 
$\langle T_c\big(
\phi^*_j(x_1') \phi_i(x_1) \phi^*_l(x_2') \phi_k(x_2) \big) \rangle$
with $T_c$ being the operator which orders the field operators
along the contour. Then, the Wick theorem tells us that 
\ba
\label{Wick}
\langle T_c\big( \phi^*_j(x_1') \phi_i(x_1) 
\phi^*_l(x_2') \phi_k(x_2) \big) \rangle 
&=& \langle T_c\big( \phi^*_j(x_1') \phi_i(x_1) \big) \rangle
\langle T_c\big( \phi^*_l(x_2') \phi_k(x_2) \big) \rangle
\\[2mm] \nonumber
&+& 
\langle T_c\big( \phi^*_j(x_1') \phi_k(x_2) \big) \rangle
\langle T_c\big( \phi^*_l(x_2') \phi_j(x_1) \big) \rangle \;,
\ea
when the field $\phi_i(x)$ is free. The Wick decomposition of
expectation value of path ordered product of field operators is 
carefully discussed in Appendix A in \cite{Danielewicz84}. We only 
mention here that there are some limitations on the decomposition 
if there are non-trivial correlations in the initial state of interest. 
However, we are not going to consider such states.

Keeping in mind, how the time arguments of $x_1, x_1', x_2, x_2'$ are 
located on the contour in Fig.~\ref{fig-contour}, the result (\ref{Wick}) 
is rewritten as 
\ba
\label{Wick-on-contour}
\langle \phi^*_j(x_1') \phi_i(x_1)
\phi^*_l(x_2') \phi_k(x_2) \big) \rangle
&=& \langle \phi^*_j(x_1') \phi_i(x_1) \rangle
    \langle \phi^*_l(x_2') \phi_k(x_2) \rangle
\\[2mm] \nonumber
&+&
\langle  \phi^*_j(x_1') \phi_k(x_2)  \rangle
\langle  \phi_j(x_1) \phi^*_l(x_2') \rangle \;,
\ea
and consequently,
\be
\label{W-final}
W_{ijkl}(x_1,x_1',x_2,x_2') =
\langle  \phi^*_j(x_1') \phi_k(x_2)  \rangle
\langle  \phi_j(x_1) \phi^*_l(x_2') \rangle 
= i\Delta^<_{kj}(x_2,x_1') \; i\Delta^>_{jl}(x_1,x_2') 
\;.
\ee

Substituting the result (\ref{W-final}) into Eq.~(\ref{corr-fields}), 
one finds
\ba
\label{corr-77}
\langle \delta {\cal F}_{ij}(X_1,{\bf p}_1)\:
\delta {\cal F}_{kl}(X_2,{\bf p}_2) \rangle 
&=&
4E_{{\bf p}_1} E_{{\bf p}_2} \int \frac{dp^0_1}{2\pi}\, \Theta(p_1^0)
\int \frac{dp^0_1}{2\pi} \, \Theta(p_2^0)
\int d^4u_1 \int d^4u_1 \: e^{i(p_1 u_1 + p_2 u_2)}
\\[2mm] \nonumber
&\times& 
i\Delta^<_{kj}(X_2+\frac{u_2}{2},X_1-\frac{u_1}{2}) \; 
i\Delta^>_{il}(X_1+\frac{u_1}{2},X_2-\frac{u_2}{2})
\\[2mm] \nonumber
&=&
4E_{{\bf p}_1} E_{{\bf p}_2} \int \frac{dp^0_1}{2\pi}\, \Theta(p_1^0)
\int \frac{dp^0_1}{2\pi} \, \Theta(p_2^0)
\int d^4u_1 \int d^4u_1
\\[2mm] \nonumber
&\times& 
\int \frac{d^4k_1}{(2\pi)^4}
\int \frac{d^4k_2}{(2\pi)^4}
\; e^{i(p_1 u_1 + p_2 u_2 - k_1\tilde u_1 - k_2\tilde u_2)}
i\Delta^<_{kj}(\tilde X_1,k_1)
i\Delta^>_{il}(\tilde X_2,k_2) \;,
\ea
where
\ban
\tilde X_1 \equiv \frac{1}{2}(X_1+X_2) - \frac{1}{4}(u_1 + u_2) 
\;,\;\;\;\;\;
\tilde u_1 \equiv X_2 - X_1 + \frac{1}{2}(u_1 + u_2) \;,
\\[2mm]
\tilde X_2 \equiv \frac{1}{2}(X_1+X_2) + \frac{1}{4}(u_1 + u_2)
\;,\;\;\;\;\;
\tilde u_2 \equiv X_1 - X_2 + \frac{1}{2}(u_1 + u_2) \;.
\ean

And now we adopt the assumption which is crucial for our further
considerations. Namely, we assume that the system under consideration
is on average homogeneous and stationary. Therefore, the Wigner
transformed Green's functions and the average distribution functions
are independent of space-time variable $X, X_1, X_2, \tilde X_1$ or
$\tilde X_2$, respectively. We also assume that the average
distrubtion function has the structure 
$\langle {\cal F}_{ij}(X,{\bf p}) \rangle
= \delta^{ij} n({\bf p})$. Then, the formulas (\ref{<-f}, \ref{>-f})
get the form
\ba
\label{<-f-eq}
i\Delta^<_{ij}(X,p) = \frac{\pi}{E_{\bf p}} \delta (E_{\bf p}-p^0)
\delta^{ij} n({\bf p})
+ \frac{\pi}{E_{\bf p}} \delta (E_{\bf p}+p^0)
\delta^{ij}
\big[ \bar n(-{\bf p}) + 1 \big] \;,
\\[2mm]
\label{>-f-eq}
i\Delta^>_{ij}(X,p) =
\frac{\pi}{E_{\bf p}} \delta (E_{\bf p}-p^0)
\delta^{ij}
\big[n({\bf p}) + 1\big]
+ \frac{\pi}{E_{\bf p}} \delta (E_{\bf p}+p^0)
\delta^{ij} \bar n(-{\bf p}) \;.
\ea

Substituting the Green's functions (\ref{<-f-eq}, \ref{>-f-eq})
into Eq.~(\ref{corr-77}), the integrals over $p^0_1$,  $p^0_2$,
$u_1$ and $u_2$ can be trivially performed and one finds
\ban
\langle \delta {\cal F}_{ij}(X_1,{\bf p}_1)\:
\delta {\cal F}_{kl}(X_2,{\bf p}_2) \rangle 
&=&
\delta^{il} \delta^{jk}
\int \frac{d^4k_1}{(2\pi)^4}
\int \frac{d^4k_2}{(2\pi)^4}
\frac{E_{{\bf p}_1} E_{{\bf p}_2}}{E_{{\bf k}_1}E_{{\bf k}_2}}
\; e^{i(k_1-k_2)(X_1-X_2)}
\\[2mm] \nonumber
&\times&
(2\pi)^4 \delta^{(4)}\Big(p_1-\frac{k_1}{2} -\frac{k_2}{2}\Big)
(2\pi)^4 \delta^{(4)}\Big(p_2-\frac{k_1}{2} -\frac{k_2}{2}\Big)
n({\bf k}_1) \big[1 + n({\bf k}_2)\big] \;.
\ean
Using the variables $Q \equiv (k_1+k_2)/2$ and $q \equiv k_1 -k_2$,
we finally obtain the main result of the Appendix
\ba
\label{corr-final-quant}
\langle \delta {\cal F}_{ij}(X_1,{\bf p}_1)\:
\delta {\cal F}_{kl}(X_2,{\bf p}_2) \rangle 
&=&
\delta^{il} \delta^{jk}
(2\pi)^3 \delta^{(3)}({\bf p}_1-{\bf p}_2)
\\[2mm] \nonumber
&\times&
\int \frac{d^3q}{(2\pi)^3}
\frac{E_{{\bf p}_1} E_{{\bf p}_2}}
{E_{{\bf p}_1 + {\bf q}/2}E_{{\bf p}_1-{\bf q}/2}}
\; e^{iq(X_1-X_2)}
n({\bf p}_1+{\bf q}/2) \big[1 + n({\bf p}_1-{\bf q}/2) \big] \;,
\ea
where $q^0 \equiv E_{{\bf p}_1 + {\bf q}/2} - E_{{\bf p}_1 - {\bf q}/2}$. 
Another derivation of the formula analogous to (\ref{corr-final-quant})  
for particles with no internal degrees of freedom or for particles
with spin can be found in \cite{Tsytovich89}. In our opinion, however, 
the decomposition, which corresponds to our Eq.~(\ref{Wick-on-contour}),
is not very convincing as obtained in \cite{Tsytovich89}. Just to
justify this step of derivation, we have referred to the Keldysh-Schwinger
technique. 

One observes that the main contribution to the integral over ${\bf q}$
in Eq.~(\ref{corr-final-quant}) comes from such ${\bf q}$ that 
$|{\bf q}| \le 1/|{\bf X}_1-{\bf X}_2|$. If the characteristic (momentum) 
scale at which the distribution function $n({\bf p})$ changes sizably 
(for the equilibrium gas of massless particles the scale is given 
by the gas temperature ($T$)) is much bigger than $1/|{\bf X}_1-{\bf X}_2|$
(for the equilibrium gas we require $1 \ll |{\bf X}_1-{\bf X}_2|T$)), the 
function under the integral can be approximated assuming that 
$|{\bf q}| \ll |{\bf p}_1|$. Then, $q^0 = {\bf v}_1{\bf q}$ and one finds
the classical correlation function
\ba
\label{corr-final-class}
\langle \delta {\cal F}_{ij}(X_1,{\bf p}_1)\:
\delta {\cal F}_{kl}(X_2,{\bf p}_2) \rangle 
&=&
\delta^{il} \delta^{jk}
(2\pi)^3 \delta^{(3)}({\bf p}_1-{\bf p}_2)\;
\delta^{(3)}({\bf X}_1-{\bf X}_2- {\bf v}_1(t_1-t_2)) \;
n({\bf p}_1) \;,
\ea
where we have additionally assumed that populations of the system's 
modes are small ($ n({\bf p}_1) \ll 1$). Eq.~(\ref{corr-final-class}),
as well as Eq.~(\ref{corr-final-quant}), is valid for both equilibrium
and nonequilibrium systems.



\begin{thebibliography}{99}

\bibitem{Mrowczynski:2005ki}
St.~Mr\'owczy\'nski,
Acta Phys.\ Polon.\  B {\bf 37}, 427 (2006).

\bibitem{Asakawa:2006jn}
M.~Asakawa, S.~A.~Bass and B.~Muller,
Prog.\ Theor.\ Phys.\  {\bf 116}, 725 (2007).

\bibitem{Arnold:2005vb}
P.~Arnold, G.~D.~Moore and L.~G.~Yaffe,
Phys.\ Rev.\  D {\bf 72}, 054003 (2005).

\bibitem{Arnold:2005ef}
P.~Arnold and G.~D.~Moore,
Phys.\ Rev.\  D {\bf 73}, 025006 (2006).

\bibitem{Rebhan:2005re}
A.~Rebhan, P.~Romatschke and M.~Strickland,
JHEP {\bf 0509}, 041 (2005).

\bibitem{Dumitru:2006pz}
A.~Dumitru, Y.~Nara and M.~Strickland,
Phys.\ Rev.\  D {\bf 75}, 025016 (2007).

\bibitem{Romatschke:2006nk}
P.~Romatschke and R.~Venugopalan,
Phys.\ Rev.\  D {\bf 74}, 045011 (2006).

\bibitem{Bodeker:2007fw}
D.~Bodeker and K.~Rummukainen,
JHEP {\bf 0707}, 022 (2007).

\bibitem{Berges:2007re}
J.~Berges, S.~Scheffler and D.~Sexty,
Phys.\ Rev.\  D {\bf 77}, 034504 (2008).
 
\bibitem{Akh75} A.I.~Akhiezer, I.A.~Akhiezer, R.V.~Polovin, A.G.~Sitenko, 
and K.N.~Stepanov, {\it Plasma Electrodynamics} (Pergamon, New York, 1975).

\bibitem{Sit82} A.G.~Sitenko, {\it Fluctuations and Non-Linear Wave 
Interactions in Plasmas}, (Pergamon, Oxford, 1982).

\bibitem{Siv85} H.D.~Sivak, Ann. Phys. (N.Y.) {\bf 159}, 351 (1985).

\bibitem{Lemoine:1995fh}
D.~Lemoine,
Phys.\ Rev.\  D {\bf 51}, 2677 (1995).

\bibitem{LP81} E.M.~Lifshitz and L.P.~Pitaevskii,
{\it Physical Kinetics} (Pergamon Press, Oxford, 1981).

\bibitem{Mrow2007}
St.~Mr\'owczy\'nski,
arXiv:0711.2003 [physics], to appear in Acta Phys. Pol. {\bf B}.

\bibitem{Litim:1999id}
D.~F.~Litim and C.~Manuel,
Nucl.\ Phys.\  B {\bf 562}, 237 (1999).

\bibitem{Litim:2001db}
D.~F.~Litim and C.~Manuel,
Phys.\ Rept.\  {\bf 364}, 451 (2002).

\bibitem{Mrowczynski:2000ed}
St.~Mr\'owczy\'nski and M.~H.~Thoma,
Phys.\ Rev.\  D {\bf 62}, 036011 (2000).

\bibitem{Thoma:1995ju}
M.~H.~Thoma,
in {\it Quark-Gluon Plasma 2}, ed. R.C. Hwa
(World Scientific, Singapore, 1995).

\bibitem{Mrowczynski:2004kv}
St.~Mr\'owczy\'nski, A.~Rebhan and M.~Strickland,
Phys.\ Rev.\  D {\bf 70}, 025004 (2004).

\bibitem{Arnold:1998cy}
P.~Arnold, D.~T.~Son and L.~G.~Yaffe,
Phys.\ Rev.\  D {\bf 59}, 105020 (1999).

\bibitem{Mrowczynski:1996vh}
St.~Mr\'owczy\'nski,
Phys.\ Lett.\  B {\bf 393}, 26 (1997).

\bibitem{Mrowczynski:1992hq}
St.~Mr\'owczy\'nski and U.~W.~Heinz,
Annals Phys.\  {\bf 229}, 1 (1994).

\bibitem{Danielewicz90}
P.~Danielewicz,
Ann. Phys. {\bf 197}, 154 (1990).

\bibitem{Danielewicz84}
P.~Danielewicz, Ann. Phys. {\bf 152}, 239 (1984).

\bibitem{Tsytovich89}
V.N.~Tsytovich,
Usp. Fiz. Nauk {\bf 159}, 335 (1989) [Sov. Phys. Usp. {\bf  32}, 911 (1989)].


\end{thebibliography}
\end{document}